\documentclass[aps,pra,twocolumn,showpacs,amsfonts,amsmath,floatfix,byrevtex]{revtex4-1}
\usepackage{graphicx}

\begin{document}

%\preprint{}

\title{Pair tunneling of two atoms out of a trap}

\author{Massimo Rontani}

\email{massimo.rontani@nano.cnr.it}
\homepage{www.nano.cnr.it}

\affiliation{CNR-NANO Research Center S3, Via Campi 213a, 41125 Modena, Italy}


%\author{}
%\thanks{}
%\affiliation{}


\date{\today}

\begin{abstract}
A simple theory for the tunneling of two cold atoms out of a trap
in the presence of an attractive contact force is developed.
Two competing decay channels, respectively for single-atom 
and bound-pair tunneling, contribute independently to the decay law
of the mean atom number in the trap.
The single-atom tunneling rate 
is obtained through the quasiparticle wave function formalism.
For pair tunneling an effective equation for the center-of-mass 
motion is derived, so the calculation of the
corresponding tunneling rate is again reduced to a 
simpler one-body problem. 
The predicted dependence of tunneling rates on the interaction strength 
qualitatively agrees with a recent measurement of the two-atom decay time
[G. Z\"urn, A. N. Wenz, S. Murmann, T. Lompe, and S. Jochim, 
arXiv:1307.5153].
\end{abstract}

\pacs{67.85.Lm, 03.75.Lm, 03.75.Ss, 74.50.+r}

\maketitle

\section{Introduction}

Pairing between two-species fermions leads to fascinating superfluid
properties of quantum systems as diverse as electrons in 
metals \cite{deGennes1999}, protons
and neutrons in nuclei \cite{Migdal1960,Bohr2008}
and neutron stars \cite{Ginzburg1964,Pines1969}, 
$^3$He atoms \cite{Legget2006},
electrons and holes in semiconductors \cite{Rontani2013}. 
In the celebrated case
of the superconductivity of metals, tunneling
spectroscopies played a major role in the 
confirmation of the theory by 
Bardeen, Cooper, and Schrieffer (BCS) \cite{BCS1957}.
Hallmark phenomena of superconductivity such as the Josephson
effect \cite{Barone1982} and the Andreev reflection \cite{Andreev1964}
were explained in terms of correlated
tunneling of two bound electrons of opposite spin (a Cooper pair).
Recently, it became possible to confine in an optical trap a few cold
$^6$Li atoms behaving as fermions of spin one-half
with unprecedented degree of control \cite{Serwane2011}. 
By properly shaping the
confinement potential in time, one may prepare exactly $N$ 
atoms in their ground state, let them tunnel out of the trap,
and measure the decay time \cite{Serwane2011,Zuern2012,Zuern2013}. 
Contrary to the case of the loosely bound
Cooper pairs, whose binding energy is fixed by the Debye frequency
of the metal, the attractive two-body interaction between the $^6$Li atoms
may be tuned through a Feshbach resonance \cite{Chin2010}.  
This allows in principle to observe the decay 
due to pair tunneling in the whole regime of interaction,
from BCS-like weakly-bound pairs to strongly bound $^6$Li
molecules undergoing Bose-Einstein 
condensation \cite{Legget2006,Regal2003,Bartenstein2004,Zwierlein2005,Giorgini2008}.

Here we focus on the basic case of two atoms
in a trap---the building block
of many-body states---and develop a simple theory 
of the decay time in the presence of an attractive contact interaction.
Within a rate-equation approach, both single-atom and
pair tunneling independently contribute to the decay of the average
number of atoms in the trap.
We compute the single-atom tunneling rate
considering the interaction of the tunneling quasiparticle with
the atom left in the trap \cite{Rontani2012}.
To obtain the pair tunneling rate, we derive an effective one-body
Schr\"odinger equation for the center-of-mass
motion and apply the semiclassical
Wentzel-Kramers-Brillouin (WKB)
formula [cf.~Eq.~\eqref{Zuernformula2}]. 

We consider the recent measurement of the decay time reported 
by the Heidelberg group in Ref.~\onlinecite{Zuern2013}, ignoring 
complications of the actual experiment that might be important for 
a quantitative comparison with the theory.
These include the effect of the trap anharmonicity on the
two-body wave function as well as the slight difference
between the magnetic moments of the two atomic species. 
Nevertheless, the dependence of the decay time
on the interaction strength that we predict
qualitatively compares with the measured trend, as shown
in Fig.~\ref{taus_omega0_25}.
The Heidelberg experiment could not single out unambiguously 
the contribution of pair tunneling          
to the decay time, due to the large uncertainty in detecting
survivor atoms in the trap for increasing attractive interactions.
Our theory highlights that the signatures of pair tunneling are
within reach of future experiments at moderate regimes of interaction. 

Stimulated by experimental advances 
\cite{Dudarev2007,Cheinet2008,Serwane2011,Zuern2012}, 
a fast-growing theoretical
literature has been focusing on different aspects of tunneling in few-atom 
traps. One theme regards multiparticle
noninteracting states and the so called Fermi-Bose 
duality \cite{delCampo2006,delCampo2011,Calderon2011,Rontani2012,Sokolovski2012,Longhi2012,Pons2012,Georgiou2012}. 
The latter refers to the feature of one-dimensional systems that
noninteracting fermions own the same observable properties as
interacting bosons when their inter-particle contact forces acquire infinite 
strength \cite{Girardeau1960,Cheon1999}.
A second theme is the tunneling dynamics in double or
multiple wells in the presence of a repulsive interaction, which drives
the competition between Josephson-like oscillations and
two-particle correlated 
tunneling \cite{Zollner2008,Streltsov2011,Chatterjee2012,Hunn2013,Bugnion2013}.
A few works have addressed the full quantum mechanical
time evolution of two interacting atoms,
tunneling out of a trap into 
free space, limitedly to repulsive interactions
and idealized geometries \cite{Lode2009,Kim2011b,Lode2012,Hunn2013}.
Within time-dependent perturbation theory \cite{Bardeen1961}, 
Ref.~\onlinecite{Rontani2012}
has computed the quasiparticle decay time of two $^6$Li atoms,
either in their ground state
with strong repulsive interactions or in the `super-Tonks-Girardeau'
excited state \cite{Girardeau2010,Girardeau2010b},  
as measured in Ref.~\onlinecite{Zuern2012}. 
In the experiment the two 
energy branches were accessed 
by scanning the Feshbach resonance through the Fermi-Bose duality point.
The influence of ferromagnetic
spin correlations on tunneling has been investigated 
in Ref.~\onlinecite{Bugnion2013b} using Fermi golden rule. 
We are aware of only one theoretical study of two particles 
attracting each other 
that tunnel out of a trap \cite{Taniguchi2011},
although limited to long-range Coulomb interactions.

Our approach based on rate equations is in principle subject to
two types of limitations: 
(i) It gives an approximate treatment 
of tunneling at the single-particle level, providing
an exponential decay law. 
The latter deviates from the exact behavior 
(see e.g.~\cite{Razavy2003}) both at 
short (Zeno effect) \cite{Wilkinson1997} and long times \cite{Rothe2006}.
(ii) It neglects higher-order correlations between single-atom
and pair tunneling channels. Such correlations may be taken into
account when considering the full time evolution of the interacting 
wave function \cite{Lode2009,Kim2011b,Lode2012}.
There is presently no indication that issues (i) and (ii) are
relevant for the class of experiments we analyze here
\cite{Zuern2012,Rontani2012,Zuern2013}.

The structure of this Article is the following:
The model Hamiltonian is presented in Sec.~\ref{s:model},
the decay law is derived in sec.~\ref{s:decay},
the tunneling rates are obtained in Secs.~\ref{s:singleatom}
and \ref{s:pairatom},
the numerical results are discussed and compared
with the Heidelberg experiment in Sec.~\ref{s:result}.

\section{Two fermions in a trap}\label{s:model}

\begin{figure}
\begin{center}
\includegraphics[width=8.0cm, angle=0]{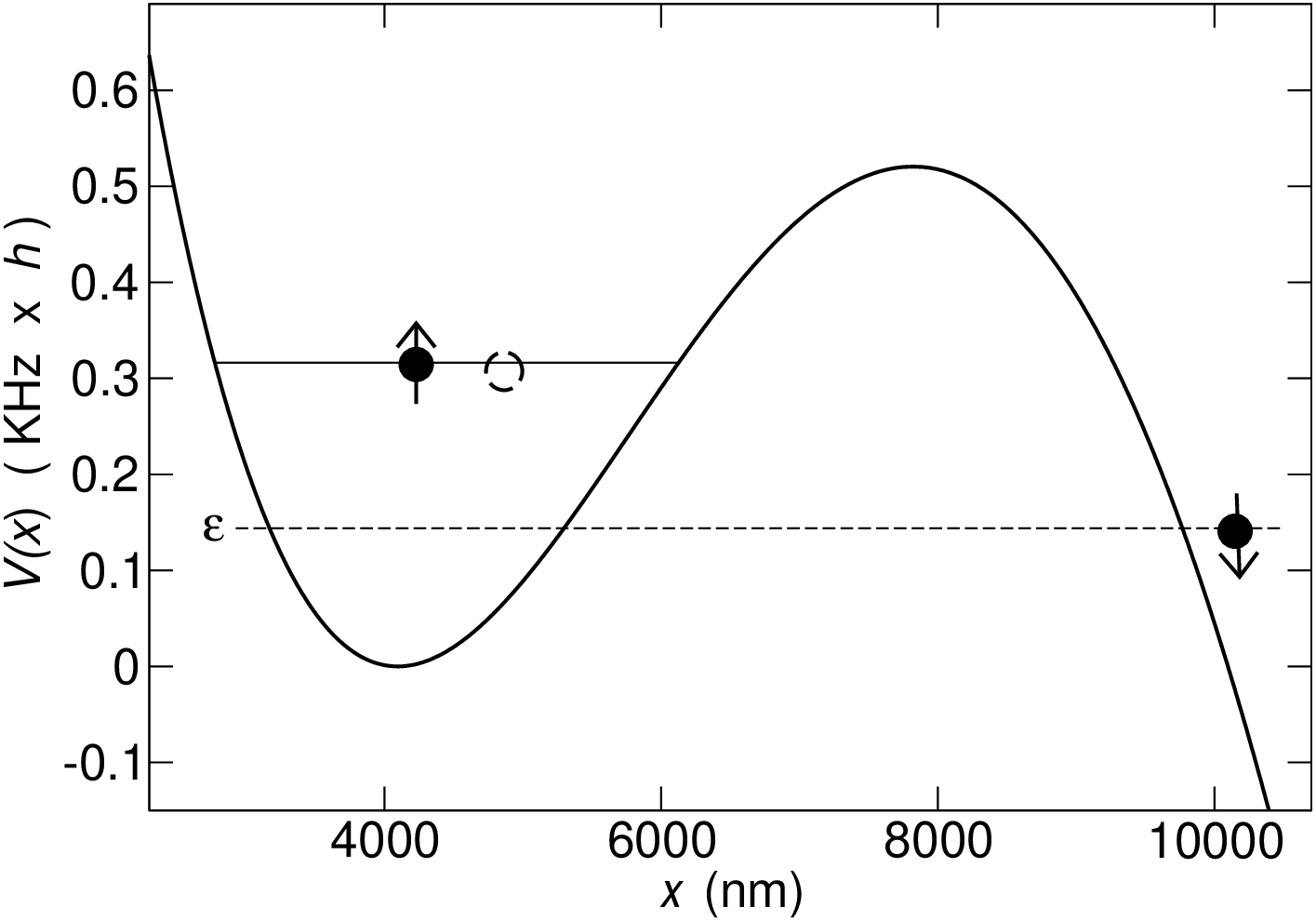}
\end{center}
\caption{ 
Confinement potential $V(x)$ vs $x$ as in
Ref.~\onlinecite{Zuern2013} with $p = 0.6338$
and $c_{B\left|\text{state}\right>}=1$.  
The frequency $\omega_{\text{WKB}}$
of the single bound state (solid thin line)
computed through the WKB approximation
is $\omega_{\text{WKB}}=$
316.3 Hz $\times$ $2\pi$ (the trap bottom is the frequency zero).
The tunneling energy for single-atom
escape is $\varepsilon$. 
Note that for attractive interactions ($g<0$)
one has $\varepsilon < \hbar \omega_{\text{WKB}}$
hence single-atom resonant tunneling is suppressed
when $\varepsilon < 0$.
The circles with arrows schematize the atoms in the
$\left|\uparrow\downarrow\right>$ final configuration.
\label{frame}}
\end{figure}

In the combined optical and magnetic potential 
illustrated in Fig.~\ref{frame}, which is quasi
one-dimensional due to the strong transverse confinement,
two $^6$Li atoms behave as fermions of spin one-half and
interact through an attractive tunable contact potential,
$g\,\delta(x_1-x_2)$, with $g<0$ (for the regime of strong repulsion 
see Ref.~\onlinecite{Wang2012b}).
A finite and smooth tunnel barrier allows atoms to escape from the trap
into the unbound region at large positive values of $x$.
The Hamiltonian is
\begin{equation}
H = -\frac{\hbar^2}{2m}\sum_{i=1}^2
\left[\frac{d^2}{dx_i^2}+V(x_i)\right]+g\delta(x_1-x_2),
\label{eq:Hx1x2}
\end{equation}
with $m$ being the mass and $V(x)$
an effective potential. The exact functional
form of $V(x)$ reproduces the setup of Ref.~\onlinecite{Zuern2013}
(cf.~Supplemental Material)  
with the optical trap depth parameter $p=0.6338$
and $c_{B\left|\text{state}\right>}=1$.
The latter condition is equivalent to neglect the weak dependence of
the atom magnetic moment (and hence the
potential profile) on the atom spin.

The trap is approximately parabolic at low energy,
the trap bottom being the energy zero. 
The bound trap eigenstates $\phi_n(x)$ are therefore
eigenstates of the one-dimensional harmonic oscillator (HO),
with energy $\varepsilon_n = \hbar\omega_0(n+1/2)$ and $n=0,1,2,\ldots$
The characteristic HO length is $\ell_{\text{HO}}=(\hbar/m\omega_0)^{1/2}$.
According to a WKB calculation, $V(x)$ supports a single bound state
(cf.~the solid thin line in Fig.~\ref{frame}),
with $\omega_{\text{WKB}}=$ 316.3 Hz $\times$ $2\pi$. 

Here only the $\left|\uparrow\downarrow\right>$
configuration of 
distinguishable fermions is considered, with the two atoms of opposite spin
being paired in their ground state. Therefore, the orbital part of the
$\left|\uparrow\downarrow\right>$ wave function is bosoniclike,
$\psi_{\uparrow\downarrow}(x_1,x_2) =
\psi_{\uparrow\downarrow}(x_2,x_1)$.
The notation used throughout this Article is consistent with that of
Ref.~\onlinecite{Rontani2012}.

\section{Decay law}\label{s:decay}

In the experiment of Ref.~\onlinecite{Zuern2013} two  atoms
are initially prepared in the ground state of the optical trap
where the coupling constant $g$ is set by a Feshbach resonance
tuning the magnetic offset field to a fixed value. 
A magnetic field gradient is then applied along $x$
for a given hold time $t$,
whose effect is to add a linear term to the confinement potential.
The net result, shown in Fig.~\ref{frame}, is to lower the 
potential barrier allowing for atoms to tunnel out of the trap.
The measurement cycle ends by ramping the potential barrier back up
and then counting the $N(t)$ survivor atoms left in the trap.
Averaging over many measurement cycles 
provides the probabilities $P_2(t)$, $P_1(t)$, and $P_0(t)$ 
of finding respectively
two, one, and zero atoms in the trap after the hold time $t$.
From the conservation of probability one has $P_2(t)+P_1(t)+P_0(t)=1$
at all times, therefore only two quantities are
independent, say $P_2(t)$ and $P_1(t)$. 
The measured mean atom number in the trap is
$\left<N(t)\right>=2P_2(t)+P_1(t)$.

In this section we derive the decay law of 
$\left<N(t)\right>$ 
on the basis of simple rate equations,
recalling the treatment of Ref.~\onlinecite{Zuern2013}
for the sake of clarity.
The decay is due to the combined effect of two qualitatively
different tunneling processes,
either the tunneling of a single atom or
the correlated escape of two bound atoms at once. Both mechanisms
may eventually empty the trap. 
We assume that the two tunneling rates, respectively 
$\gamma_s$ and $\gamma_p$ 
for single-atom and pair tunneling,
may be computed independently.

At time $t$, $P_2(t)$ may decrease due to both single-atom or pair tunneling,
so one has
\begin{equation}
\frac{dP_2(t)}{dt} = -\left(\gamma_s + \gamma_p\right)P_2(t).
\end{equation}
Here the small probability that two consecutive single-atom tunneling
events occur in the infinitesimal time interval $dt$ is neglected.
Moreover, it is assumed that 
$\gamma_s$ and $\gamma_p$ are constant in
time and add independently, as well as that the decay process is
irreversible. The decay law is then simply
\begin{equation}
P_2(t) = e^{-\left(\gamma_s+ \gamma_p\right)t}.
\label{eq:p2t}
\end{equation}
If $g=0$, there is no pair tunneling ($\gamma_p=0$) and
$\gamma_s$ is twice the rate $\gamma_{s0}$ for the decay
of a single atom in the trap, 
$\gamma_s=2\gamma_{s0}$ \cite{Kim2011b,Rontani2012,Zuern2013}. 

The variation in time of $P_1(t)$ is more complex,
\begin{equation}
\frac{dP_1(t)}{dt} = F_{\text{in}}(t) + F_{\text{out}}(t),
\end{equation}
as $P_1(t)$ may either increase due to the one-atom decay of the state
with two atoms, $F_{\text{in}}(t)=\gamma_sP_2(t)$, or decrease due to
the decay of the one-atom state, $F_{\text{out}}(t)=-\gamma_{s0}P_1(t)$, so 
\begin{equation}
\frac{dP_1(t)}{dt} = \gamma_s e^{-\left(\gamma_s+ \gamma_p\right)t}  
-\gamma_{s0}P_1(t). 
\label{eq:P1t}
\end{equation}
Solving Eq.~(\ref{eq:P1t}) with the initial condition that there are
two atoms in the trap [$P_2(0)=1$, $P_1(0)=P_0(0)=0$], the decay law is
\begin{equation}
P_1(t)=\frac{\gamma_s}{\gamma_s+\gamma_p-\gamma_{s0}}
\left[e^{-\gamma_{s0}t} - e^{-\left(\gamma_s+ \gamma_p\right)t} \right]. 
\label{eq:p1t}
\end{equation}
Therefore, the decay law for the mean particle number in the trap is:
\begin{eqnarray}
\left<N(t)\right> &=&
\left[ 2 - \frac{\gamma_s}{\gamma_s+\gamma_p-\gamma_{s0}}\right]
e^{-\left(\gamma_s+ \gamma_p\right)t}\nonumber\\ &+& 
\frac{\gamma_s}{\gamma_s+\gamma_p-\gamma_{s0}}e^{-\gamma_{s0}t} .
\label{eq:Nt}
\end{eqnarray}

In the noninteracting case one has
\begin{equation}
P_2(t) = e^{-2\gamma_{s0} t},
\label{eq:p2t0}
\end{equation}
\begin{equation}
P_1(t)=2e^{-\gamma_{s0} t}\left[ 1 - e^{-\gamma_{s0} t} \right],
\end{equation}
\begin{equation}
P_0(t)=\left[ 1 - e^{-\gamma_{s0} t} \right]^2,
\end{equation}
and it is easy to show that
\begin{equation}
P_1 = \left<N\right>-\frac{1}{2}\left<N\right>^2
\label{eq:P1fancy}
\end{equation}
at all times $t$, which is the black dashed parabola of Fig.~3 
in \cite{Zuern2013}.
At finite $g$ it is not possible to work out a relation similar to
(\ref{eq:P1fancy}) in closed form. 

The next two sections explain the calculation of the tunneling
rates $\gamma_s$, $\gamma_{s0}$, and $\gamma_p$ that
enter the expressions
\eqref{eq:p2t} and \eqref{eq:p1t} for $P_2(t)$ and $P_1(t)$, respectively.

\section{Single-atom tunneling}\label{s:singleatom} 

This section focuses on the elementary tunneling event of
a single atom transferred out of the trap. 
In the case there are initially two atoms in the trap, i.e., 
the tunneling transition is $N=2\rightarrow N=1$, 
the tunneling rate $\gamma_s$
is computed by means of the quasiparticle wave function 
theory developed in Ref.~\onlinecite{Rontani2012}.
Such approach fully takes into account the interaction between the
escaping atom and the companion left in the trap. 
The single-atom tunneling rate $\gamma_s$ is $1/\tau$
in the notation of \cite{Rontani2012}. 

For attractive interactions,
the relevant tunneling transition is the one between the initial trap ground
state $\Psi_0(x_1,x_2)$ and the final noninteracting configuration
$\Psi_{0,\varepsilon}(x_1,x_2)$, 
with one atom left in the lowest HO orbital $\phi_0(x)$
and the other one in the continuum state $\chi_{\varepsilon}(x)$ outside
the trap (Fig.~\ref{frame}).
If the total energy of the interacting state with two atoms in the trap
is $W_0(g)$,
from energy conservation it follows that the tunneling energy is
$\varepsilon = W_0(g) - \varepsilon_0$. 
Here the two-atom energies and wave functions are computed
in the harmonic approximation following the exact solution of 
Ref.~\onlinecite{Busch1998}. 

\begin{figure}
\begin{center}
\includegraphics[width=8.0cm, angle=0]{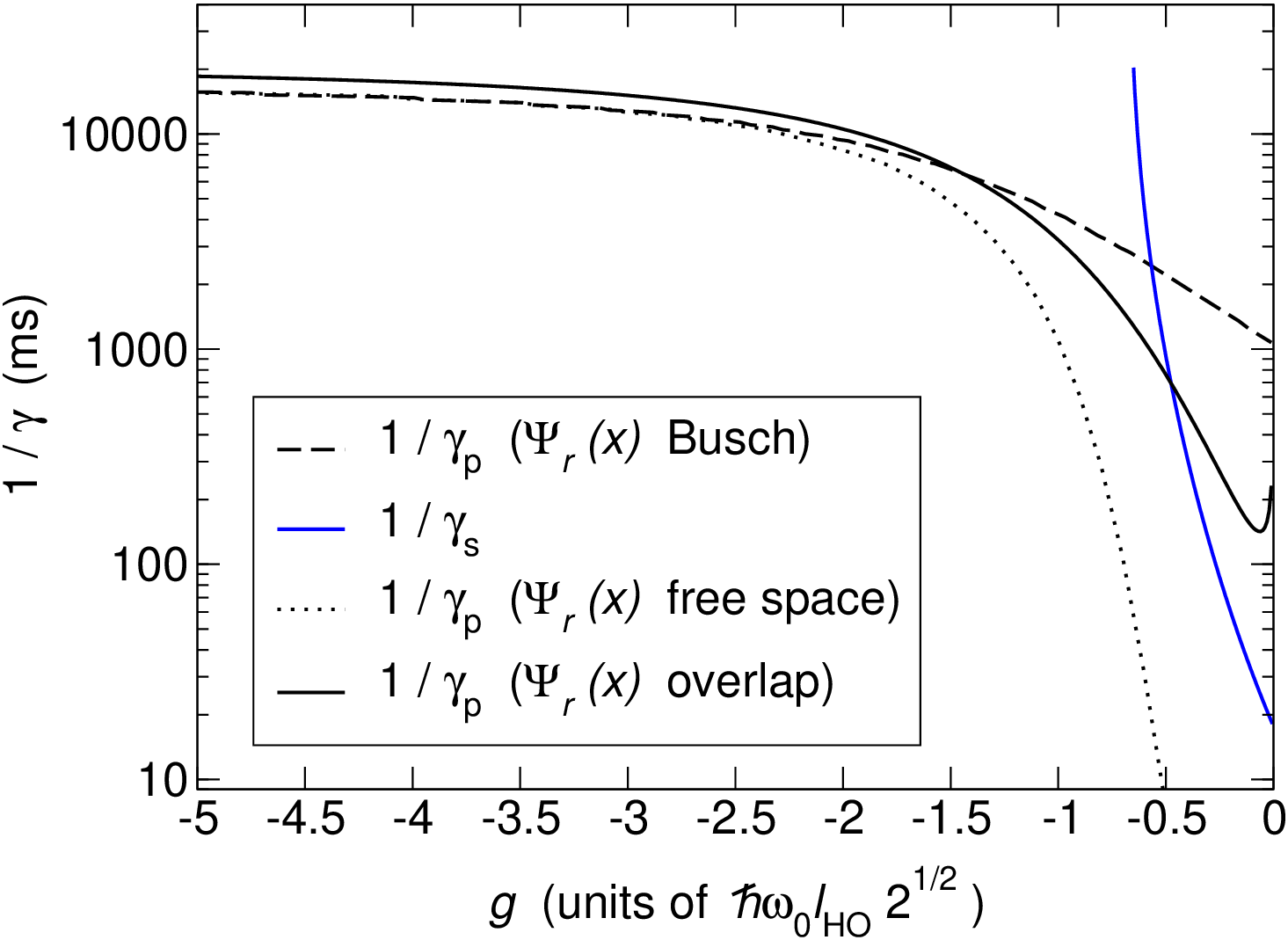}
\end{center}
\caption{(color online). 
Inverse tunneling rates $1/\gamma$ vs coupling constant $g$
for $\omega_0 = 2\omega_{\text{WKB}}=632.6$ Hz $\times$ $2\pi$.
$g$ is in units of $\hbar\omega_02^{1/2}\ell_{\text{HO}}$.
\label{omega0_632}}
\end{figure}

When the tunneling energy is lower
than the trap bottom, i.e., $\varepsilon < 0$, the
resonant tunneling process is forbidden.  
This is illustrated in Fig.~\ref{omega0_632},
where $1/\gamma_s$ is plotted vs $g$ (blue [gray] line)
for a trap with HO frequency
$\omega_0 = 2 \omega_{\text{WKB}}=632.6$ Hz $\times$ $2\pi$.
The tunneling energy
$\varepsilon$ and $\gamma_s$ decrease with 
increasing values of $\left|g\right|$, as the potential barrier 
faced by the escaping atom becomes higher and thicker.
At $g\approx -0.65$ the energy $\varepsilon$ reaches the bottom of the trap 
where the channel of single-atom resonant tunneling closes.

In general, there may be final states other than
$\Psi_{0,\varepsilon}$ allowed by energy conservation,
like the $(n=1)(\varepsilon=W_0-\varepsilon_1)$ configuration.
However, the corresponding matrix elements may be neglected 
since wave function tails drop exponentially with energy in the barrier.  

\begin{figure}
\begin{center}
\includegraphics[width=8.0cm, angle=0]{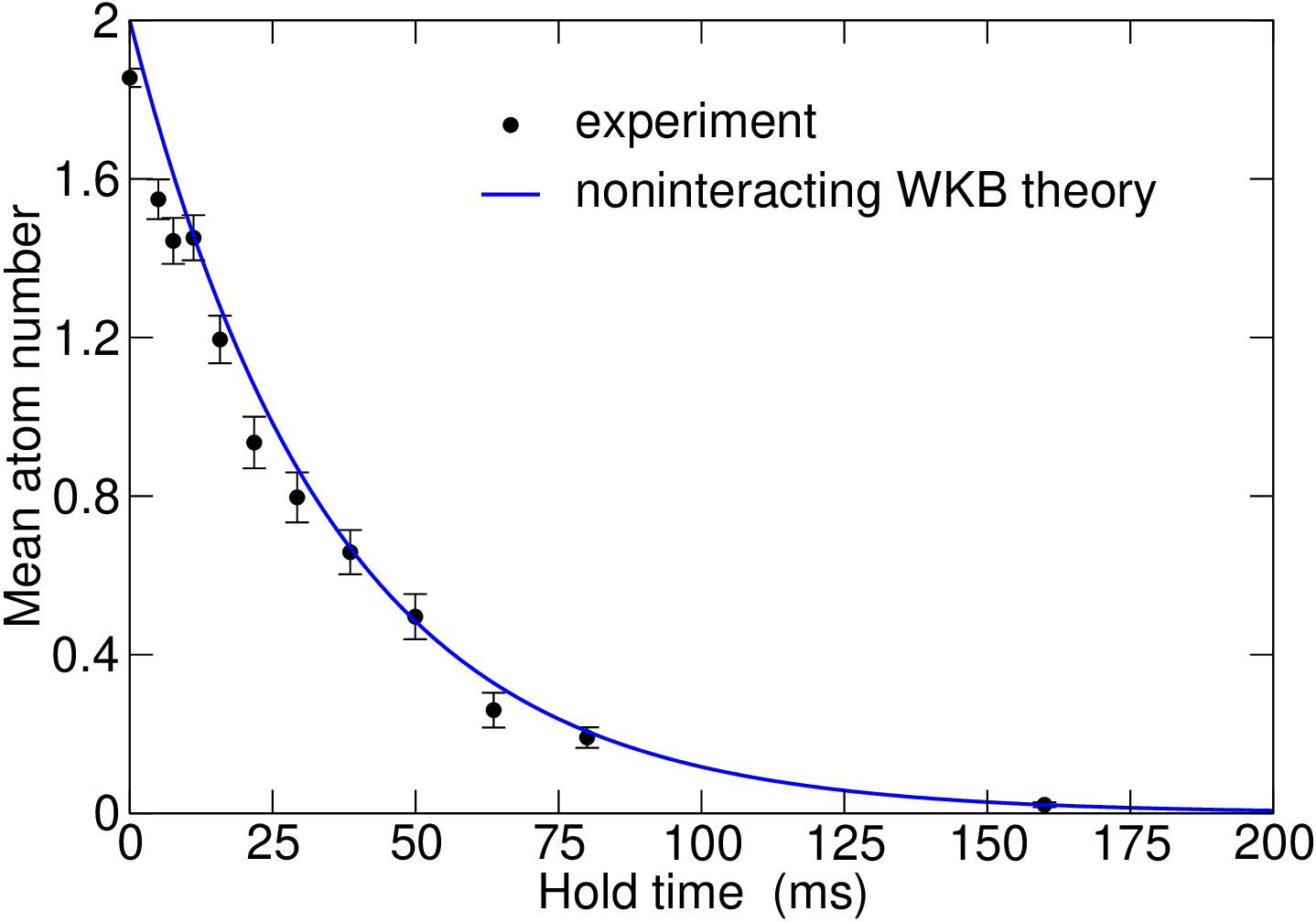}
\end{center}
\caption{(color online).  Mean atom number $\left<N(t)\right>$
vs hold time $t$. The points with their error bars are the measured 
values taken from Fig. 1 of Ref.~\onlinecite{Zuern2013} 
for $g = -0.01$ $\hbar\omega_0\ell_{\text{HO}}2^{1/2}$,
with $\omega_0=$ 632.6 Hz $\times$ $2\pi$.
The theoretical curve (continuous line) is a decaying exponential
with time constant $1/\gamma_{s0}=35.24$ ms, as obtained 
from Eq.~(\ref{Zuernformula}) with 
$\varepsilon_0 = \hbar\omega_{\text{WKB}}=316.3$ Hz $\times$ $h$,
with $\hbar\omega_{\text{WKB}}$ being the WKB energy of the 
bound state shown in Fig.~\ref{frame}.
\label{WKB}}
\end{figure}

In the case there is initially only one atom in the trap,
the tunneling rate $\gamma_{s0}$ of the 
transition $N=1\rightarrow N=0$ 
is given by the WKB formula \cite{Zuern2012}
\begin{equation}
\gamma_{s0} =
\frac{\varepsilon_0}{2\pi}\exp{\left(-2\!\!
\int_{x_a}^{x_b}\!\!\!k(x)dx\right)},
\label{Zuernformula}
\end{equation}
with $x_a<x_b$ being the classical turning points and
$k(x)=[(2m/\hbar^2)\left|\varepsilon_0-V(x)\right|]^{1/2}$.
Here it is assumed that the atom in the trap occupies the
lowest HO orbital $\phi_0(x)$.

In the case there are two noninteracting atoms 
one has $\gamma_s = 2\gamma_{s0}$
and all decay laws assume a simple form, as shown in 
Sec.~\ref{s:decay}. In particular, the mean atom number 
$\left<N(t)\right>$ is given by
\begin{displaymath}
\left<N(t)\right> =
2e^{-\gamma_{s0}t} ,
\end{displaymath}
as obtained from Eq.~(\ref{eq:Nt}) with $\gamma_p = 0$.
Figure \ref{WKB} compares such theoretical curve (continuous line),
computed for $\varepsilon_0 = \hbar \omega_{\text{WKB}}=316.3$ Hz 
$\times$ $h$,
with the experimental data \cite{Zuern2013} (points) obtained for an
almost negligible value of the coupling constant, $g = -0.01$
$\hbar\omega_0(2\hbar/m\omega_0)^{1/2}$ 
($\omega_0 = 2\omega_{\text{WKB}}=632.6$ Hz $\times$ $2\pi$), 
nicely showing the exponential decay whose 
time constant is given by the WKB prediction.

\section{Pair tunneling}\label{s:pairatom} 

In order to derive the pair tunneling rate $\gamma_p$,
we rewrite the full Hamiltonian 
(\ref{eq:Hx1x2}) in center-of-mass and relative-motion 
coordinates, 
\begin{equation}
H = -\frac{\hbar^2}{m}
\frac{d^2}{dx^2}
+g\delta(x)
-\frac{\hbar^2}{4m}
\frac{d^2}{dX^2}
+V(X+\frac{x}{2})
+V(X-\frac{x}{2}),
\label{eq:HxX}
\end{equation}
with $X=(x_1+x_2)/2$ and $x=x_1-x_2$.
The time-independent Schr\"odinger equation reads
\begin{equation}
H\Psi(X,x)=W\Psi(X,x),
\label{eq:HE}
\end{equation}
with $\Psi(X,x)$ being the two-atom ground state over the whole
space (not to be confused with Bardeen's solution $\Psi_0(x_1,x_2)$ of 
section \ref{s:singleatom}, which is the ground state in the trap region 
and vanishes outside the potential barrier \cite{Rontani2012}).
In the following we are concerned with solutions of the
eigenvalue problem (\ref{eq:HE}) such that the two
atoms form a bound state---a pair---both inside and outside the trap.

\begin{figure}
\begin{center}
\includegraphics[width=8.0cm, angle=0]{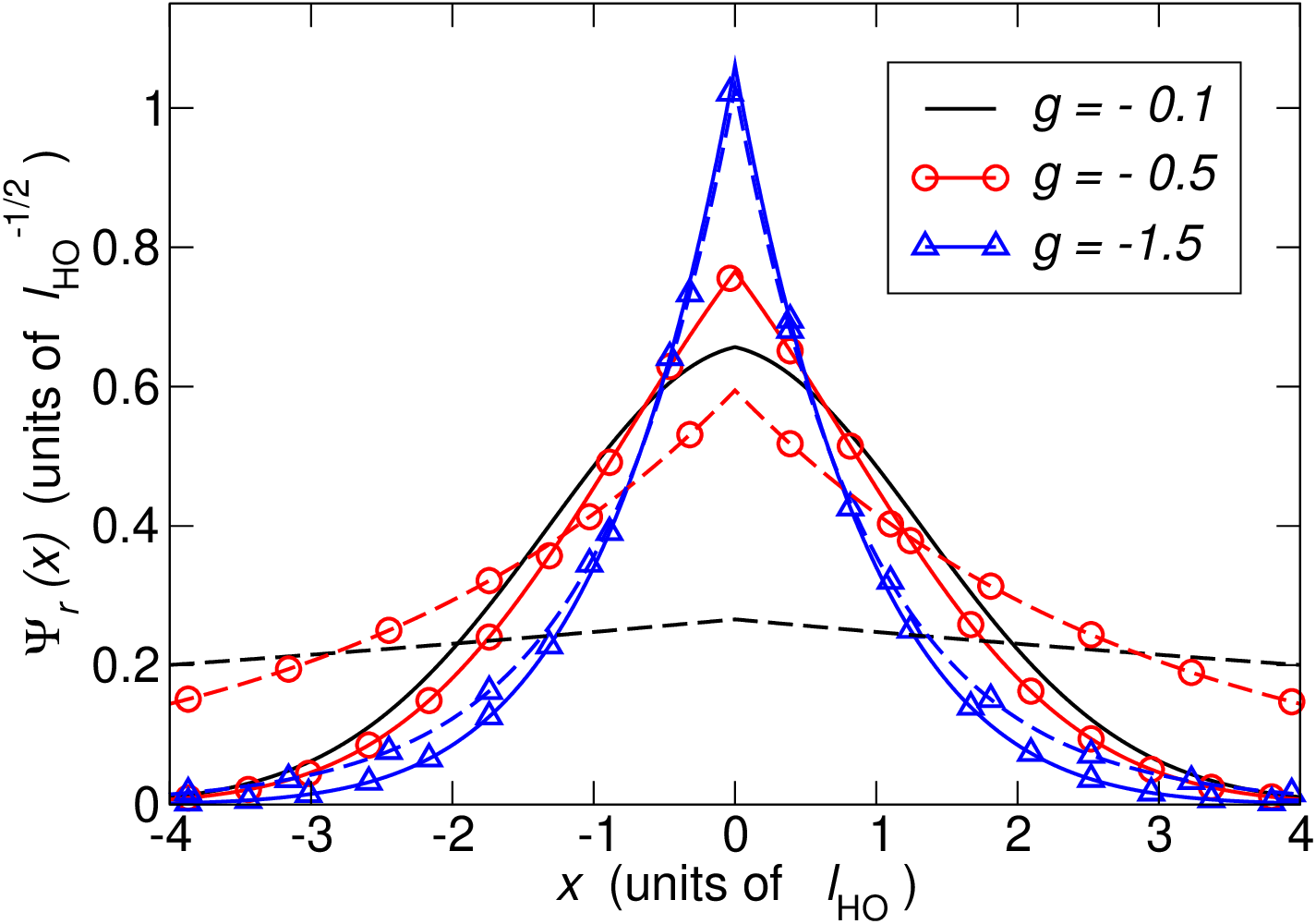}
\end{center}
\caption{(color online).
Relative-motion wave function $\Psi_r(x)$ vs $x$ 
for different values of $g$.
The solid curves represent the ground states in 
a harmonic trap \cite{Busch1998},
whereas the dashed lines show the corresponding bound states in free space. 
The length unit is $\ell_{\text{HO}}=(\hbar/m\omega_0)^{1/2}$
and $g$ is expressed in units of $\hbar\omega_0\ell_{\text{HO}}
2^{1/2}$.
\label{wf_r}}
\end{figure}

Close to the bottom of the trap the anharmonic terms of the potential
are negligible, hence
$V(X+x/2) + V(X-x/2)\approx m\omega_0^2X^2 + m\omega_0^2x^2/4$
and the Hamiltonian (\ref{eq:HxX}) becomes separable with respect to
coordinates $x$ and $X$,
the wave function being
$\Psi(X,x)=\Psi_{\text{CM}}(X)\Psi_r(x)$. In this region
one may use the exact solution for $\Psi_r(x)$ \cite{Rontani2012,Busch1998} 
whereas $\Psi_{\text{CM}}(X)$ is just a Gaussian. 
We write the total energy as $W_0=E_{\text{trap}}+\varepsilon_0$,
with only the relative-motion energy $E_{\text{trap}}$ depending on $g$.
Figure \ref{wf_r} shows $\Psi_r(x)$ for 
different values of $g$ (solid lines). Here the energy unit 
is $\hbar\omega_0$, the length unit is $\ell_{\text{HO}}$,
and $g$ is expressed in units of $\hbar\omega_0\ell_{\text{HO}} 2^{1/2}$.

Even well outside the trap the wave function is decoupled, 
$\Psi(X,x)=\Psi_{\text{CM}}(X)\Psi_r(x)$.
In this case $\Psi_{\text{CM}}(X)$ is a continuum
state whereas $\Psi_r(x)$ is finite and normalizable, with wave function
\begin{equation}
\Psi_r(x)=\frac{\sqrt{-g}}{2^{1/4}}
\exp{( g\left|x\right|/\sqrt{2} )}
\label{eq:WFfree}
\end{equation}
and energy $E_{\text{free}}=-g^2/2$
(here $\ell_{\text{HO}}=\hbar\omega_0=1$). 
The pair wave function
$\Psi_r(x)$ is compared inside (solid lines) and outside (dashed lines) the
trap in Fig.~\ref{wf_r}. For weak attraction
($g=-0.1$ and -0.5) the size of the
pair outside the trap is much larger than inside the trap, 
the trap confinement potential squeezing $\Psi_r(x)$ and forcing it
to have Gaussian tails. 
However, for stronger attraction ($g=-1.5$), the two wave functions overlap 
almost completely.

\subsection{Effective Schr\"odinger equation from ansatz
wave function}\label{s:ansatz}

In the case the two atoms tunnel as a pair, a reasonable assumption is
that the atoms form a bound state over the whole space. 
A possible ansatz wave function is then 
\begin{equation} 
\Psi(X,x) \approx \Psi_{\text{CM}}(X)\Psi_r(x) \quad \forall\,\, X,
\label{eq:ansatz}
\end{equation} 
with $\Psi_r(x)$ being the bound state wave function in the 
relative-motion frame obtained from Busch's theory \cite{Busch1998}.

By multiplying both sides of Eq.~(\ref{eq:HE}) for $\Psi^*_r(x)$,
using Eqs.~(\ref{eq:HxX}), (\ref{eq:ansatz}), and integrating over $x$, 
one obtains an effective Schr\"odinger equation for the 
center-of-mass motion:
\begin{equation} 
-\frac{\hbar^2}{4m}\frac{d^2 \Psi_{\text{CM}}(X)}{dX^2}
+V_{\text{CM}}(X) \Psi_{\text{CM}}(X) =
\varepsilon_{\text{CM}} \Psi_{\text{CM}}(X).
\label{eq:CM}
\end{equation} 
Here the effective potential $V_{\text{CM}}(X)$ is defined as
\begin{eqnarray} 
V_{\text{CM}}(X) & = & \int^{\infty}_{-\infty}
\!\!\!dx\left|\Psi_r(x)\right|^2
\nonumber \\
&\times & \left[ V(X+\frac{x}{2}) +V(X-\frac{x}{2})- 
\frac{m}{4}\omega_0^2x^2\right] ,
\label{eq:Vcm}
\end{eqnarray} 
and the center-of-mass energy $\varepsilon_{\text{CM}}$
is the total energy $W$ minus the relative-motion energy $E_{\text{trap}}$,
$\varepsilon_{\text{CM}} = W - E_{\text{trap}}$.
The energy $E_{\text{trap}}$ is the eigenvalue of the equation
\begin{equation} 
\left[ -\frac{\hbar^2}{m}
\frac{d^2}{dx^2} + 
\frac{m}{4}\omega_0^2x^2
+g\delta(x)\right]\Psi_r(x)=E_{\text{trap}}\Psi_r(x).
\end{equation} 
Equation (\ref{eq:CM}) takes into account the internal degree of
freedom of the two-atom bound state through the 
potential $V_{\text{CM}}(X)$, 
which is the original potential $V$ 
smeared by the relative-motion probability density 
$\left|\Psi_r(x)\right|^2$ appearing in (\ref{eq:Vcm}).

To shed light on the structure of $V_{\text{CM}}(X)$, 
it is useful to consider two limiting cases.
In case the potential profile is purely parabolic, $V(x)=m\omega_0^2x^2/2$,
Eq.~(\ref{eq:Vcm}) simply reduces to 
$V_{\text{CM}}(X) = m\omega_0^2X^2$ and the ansatz wave function
(\ref{eq:ansatz}) is the exact result.
The other exact limit is the case $g\rightarrow -\infty$. 
Then the pair binding energy goes to $-\infty$ and the spatial extension
of $\Psi_r(x)$ becomes negligible, hence
$\left|\Psi_r(x)\right|^2\approx \delta(x)$ and Eq.~(\ref{eq:Vcm})
becomes $V_{\text{CM}}(X) = 2V(X)$.
Equation (\ref{eq:CM}) then takes the form
\begin{equation} 
-\frac{\hbar^2}{4m}\frac{d^2 \Psi_{\text{CM}}(X)}{dX^2}
+2V(X) \Psi_{\text{CM}}(X) =
\varepsilon_{\text{CM}} \Psi_{\text{CM}}(X),
\label{eq:CMlimit}
\end{equation} 
which is the single-particle Schr\"odinger equation for a
particle of coordinate $X$ and mass $2m$ seeing the potential 2$V(X)$.

In order to compute the pair tunneling rate $\gamma_p$, it suffices
to note that the tunneling problem is reduced through (\ref{eq:CM})
to that of a single particle of mass $2m$ escaping through the effective
potential barrier defined by $V_{\text{CM}}(X)$. Then $\gamma_p$  
may be computed trough the WKB formula
\begin{equation}
\gamma_{p} =
\frac{\varepsilon_{\text{CM}}}{2\pi\hbar}\exp{\left(-2\!\!
\int_{X_a}^{X_b}\!\!\!K(X)dX\right)}.
\label{Zuernformula2}
\end{equation}
Here $X_a<X_b$ are the classical turning points for
the effective potential $V_{\text{CM}}(X)$, 
$K(X)=[(4m/\hbar^2)\left|\varepsilon_{\text{CM}}-
V_{\text{CM}}(X)\right|]^{1/2}$,
and the center-of-mass energy $\varepsilon_{\text{CM}}$ is the
WKB bound level in the trap defined by $V_{\text{CM}}(X)$.

The dashed line in Fig.~\ref{omega0_632} shows $1/\gamma_p$ obtained  
for $\omega_0 = 2\omega_{\text{WKB}}=632.6$ Hz $\times$ $2\pi$.
The inverse decay rate $1/\gamma_p$ increases with $\left|g\right|$,
the smaller the pair size the lower the tunneling rate. 
For strong attraction $1/\gamma_p$ tends
to the asymptotic exact limit of a point-like particle of mass
$2m$. However, for $g\rightarrow 0$ the decay time $1/\gamma_p$ 
disturbingly tends to a finite value, whereas one would expect it
to be suppressed as the atoms become unbound. Such difficulty is due to
the approximate form (\ref{eq:ansatz}) for $\Psi(X,x)$.

Alternatively, one may replace in the ansatz (\ref{eq:ansatz}) the trap pair
wave function with the bound state in free space [Eq.~\eqref{eq:WFfree}]. 
The corresponding
values obtained for $1/\gamma_p$ are shown by the dotted curve in
Fig.~\ref{omega0_632}. Reassuringly, for strong attraction   
the dotted and dashed curves overlap, as the pair wave functions
tend to coincide. However, as $g\rightarrow 0$ the value of $1/\gamma_p$
becomes unphysically low, since the ansatz wave function now overestimates
the pair size.  

The missing piece of information in the ansatz (\ref{eq:ansatz}) 
is the link between the pair wave function inside and outside the trap.
Indeed, one expects $1/\gamma_p$ to interpolate between
the upper and lower bounds represented respectively by the dashed and dotted
curves in Fig.~\ref{omega0_632}. 
An additional physical requirement is that $1/\gamma_p\rightarrow \infty$ as 
$g\rightarrow 0$. 
In the next subsection we propose a refined effective
potential for the center-of-mass motion which
complies with the required physical features. 

\subsection{Effective center-of-mass potential from time-dependent
perturbation theory}\label{s:PT}

As a preliminary step, we recall the result 
of time-dependent first-order perturbation theory 
\cite{Rontani2012,Bardeen1961} 
for the noninteracting single-atom tunneling rate $\gamma_{s0}$. 
Such rate is given by Fermi's golden rule,
\begin{eqnarray}
\gamma_{s0}&=&\frac{2\pi}{\hbar}
\sum_{\varepsilon}
\left|M_{0 \varepsilon}\right|^2\delta (\varepsilon_0 -\varepsilon)
\nonumber\\
&=& \frac{2\pi}{\hbar}
\left|M_{0 \varepsilon_0}\right|^2 \varrho(\varepsilon_0),
\label{eq:FGR}
\end{eqnarray}
with $\varrho(\varepsilon_0)$ being the density of continuum states 
at energy $\varepsilon_0$ and 
\begin{equation}
M_{0 \varepsilon} = \int_{-\infty}^{\infty}\!\!\!dx\;\phi_0^*(x)
\left[-\frac{\hbar^2}{2m}\frac{d^2}{dx^2}+V(x)-\varepsilon\right]
\chi_{\varepsilon}(x).
\label{eq:M}
\end{equation} 
For the following development, we remark
that the practical evaluation of $\gamma_{s0}$ relies on
the WKB formula (\ref{Zuernformula}). Therefore, we associate the
matrix element (\ref{eq:M}) with
the expression (\ref{Zuernformula}) for the decay
of a particle of energy $\varepsilon$ and mass $m$ through the
potential barrier $V(x)$.

Considering now the interacting case,
the pair-tunneling matrix element ${\cal M}_{0f}$ between the initial state
$\Psi_0(X,x)$ with two atoms in the trap
and the final state $\Psi_f(X,x)$ with the pair outside 
the trap is a straightforward extension of (\ref{eq:M}):
\begin{equation}
{\cal M}_{0f} = \int_{-\infty}^{\infty}\!\!\! dX 
\int_{-\infty}^{\infty}\!\!\! dx\; \Psi_0^*(X,x)\left[ H  - W_f\right]
\Psi_f(X,x).
\label{eq:TD}
\end{equation}
Both initial and final states 
are written as
$\Psi(X,x)=\Psi_{\text{CM}}(X)\Psi_r(x)$, the relative and 
center-of-mass motions being decoupled. 
For the initial state, $\Psi_0(X,x)=\Psi^0_{\text{CM}}(X)\Psi^0_r(x)$,
the relative-motion wave function
$\Psi^0_r(x)$ is Busch's solution with energy 
$E_{\text{trap}}(g)$ \cite{Busch1998} and
$\Psi^0_{\text{CM}}(X)$ is the lowest HO state in the center-of-mass 
frame with energy $\varepsilon_0$, the total energy being 
$W_0=E_{\text{trap}}(g) + \varepsilon_0$.
For the final state, $\Psi_f(X,x)=\Psi^f_{\text{CM}}(X)\Psi^f_r(x)$,
the relative-motion wave function
$\Psi^f_r(x)$ is the 
free-space wave function (\ref{eq:WFfree}) with energy
$E_{\text{free}}=-g^2/2$
and
$\Psi_{\text{CM}}(X)$ is the continuum state with energy
$\varepsilon_{\text{CM}}$,
the total energy being
$W_f = -g^2/2 + \varepsilon_{\text{CM}}$.

The next step is to trace out the relative-motion degree of freedom in
(\ref{eq:TD}) by integrating over $x$. One obtains
\begin{eqnarray}
{\cal M}_{0f} &=& \int_{-\infty}^{\infty}\!\!\! dX\;
\Psi^{0*}_{\text{CM}}(X)\bigg[- \frac{S\hbar^2}{4m}\frac{d^2}{dX^2}
\nonumber\\
&&+\; V_{\text{CM}}(X)-S\varepsilon_{\text{CM}}\bigg]\Psi^f_{\text{CM}}(X),
\label{eq:superM}
\end{eqnarray} 
with $S$ being the overlap integral between relative-motion 
wave functions,
\begin{equation}
S = \int_{-\infty}^{\infty}\!\!\! dx\;\Psi^{0*}_r(x)\, \Psi^f_r(x),
\end{equation}
and $V_{\text{CM}}(X)$ being the effective potential for center-of-mass
motion,
\begin{eqnarray}
V_{\text{CM}}(X) & = &
\int_{-\infty}^{\infty}\!\!\! dx\;\Psi^{0*}_r(x)\, \Psi^f_r(x)
\nonumber\\
&\times & \left[V(X+x/2)+V(X-x/2)\right].
\label{eq:Vcmoverlap}
\end{eqnarray}
By inspection we see that the formula (\ref{eq:superM})
is the matrix element for the decay of a particle of mass $2m/S$ and energy
$S\varepsilon_{\text{CM}}$ through the potential barrier 
$V_{\text{CM}}(X)$. The latter effective potential
is smeared
by the overlap density $\Psi^{0*}_r(x)\, \Psi^f_r(x)$
instead of the probability density $\left|\Psi_r(x)\right|^2$
that appears in (\ref{eq:Vcm}).

\begin{figure}
\begin{center}
\includegraphics[width=8.0cm, angle=0]{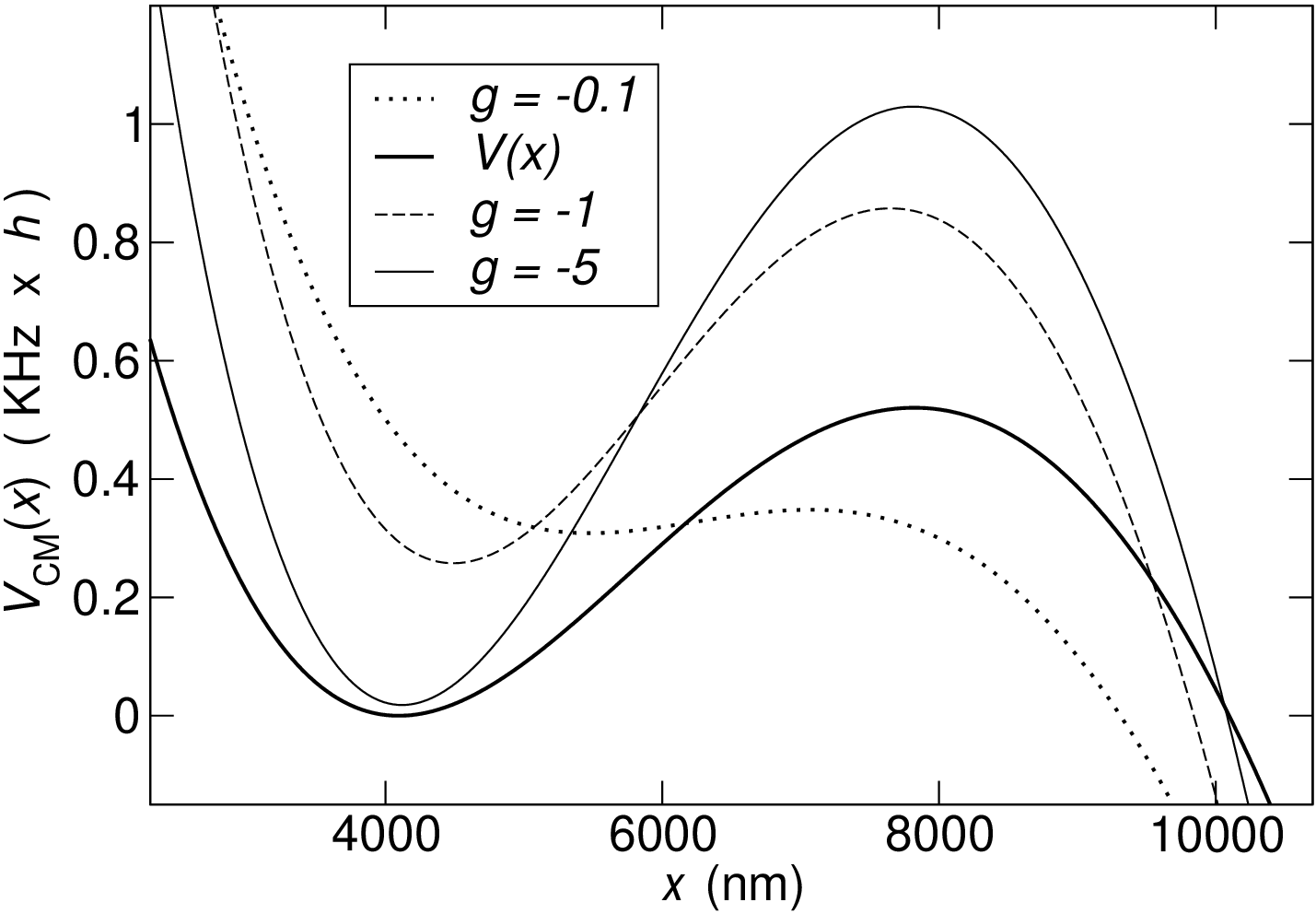}
\end{center}
\caption{ 
Effective confinement potential for the center-of-mass
motion $V_{\text{CM}}(x)$ vs $x$, as defined in Eq.~(\ref{eq:Vcmoverlap}), 
for a few values of the coupling constant $g$. Here the HO frequency 
is $\omega_0=$ 250 Hz $\times$ $2\pi$ (the trap bottom is the frequency zero).
The coupling constant
$g$ is expressed in units of $\hbar\omega_0\ell_{\text{HO}}
2^{1/2}$, with
$\ell_{\text{HO}}=(\hbar/m\omega_0)^{1/2}$.
The thick solid line is the original single-particle 
bare potential $V(x)$.
\label{V_cm}}
\end{figure}

The potential barrier induced 
by $V_{\text{CM}}(X)$ depends on the coupling constant $g$,
as shown in Fig.~\ref{V_cm}. For strong attraction, $g\rightarrow
-\infty$, the overlap density $\Psi^{0*}_r(x)\, \Psi^f_r(x)$
tends to the probability density $\left|\Psi_r(x)\right|^2$
and $S\rightarrow 1$, hence one recovers the
results of subsection \ref{s:ansatz}. In fact, the effective
particle of mass $2m/S\rightarrow 2m$ and energy 
$S\varepsilon_{\text{CM}} \rightarrow \varepsilon_{\text{CM}} $
sees a potential barrier
$V_{\text{CM}}(x)\rightarrow 2V(x)$
(compare thin and thick solid lines in Fig.~\ref{V_cm}).
However, for $g\rightarrow 0$ the effective one-particle
problem strongly deviates from that of subsection \ref{s:ansatz},
as now the particle acquires infinite mass since $S\rightarrow 0$.

The above discussion shows that $\gamma_p$ may be evaluated
by means of the WKB formula (\ref{Zuernformula2}) 
using the expression (\ref{eq:Vcmoverlap}) for $V_{\text{CM}}(X)$
and replacing $2m$ with $2m/S$ as well as 
$\varepsilon_{\text{CM}}$ with $S\varepsilon_{\text{CM}}$.
The result for $\omega_0 = 2\omega_{\text{WKB}}=632.6$ Hz $\times$ $2\pi$
is shown in Fig.~\ref{omega0_632} (black solid line).
One recovers the previous results of subsection
\ref{s:ansatz} for $g\rightarrow -\infty$, 
as all black curves tend to the same line asymptotically. 

The small discrepancy between solid and dashed / dotted lines
for large values of $\left|g\right|$ is due to the 
different method to determine $\varepsilon_{\text{CM}}$.
In fact, for the black solid curve $\varepsilon_{\text{CM}}$ is fixed by
energy conservation, $\varepsilon_{\text{CM}} = \varepsilon_0 +
E_{\text{trap}}(g)+g^2/2$, whereas for the dashed and dotted curves
$\varepsilon_{\text{CM}}$ is the WKB energy of the bound state in the
effective potential. Nevertheless, in the limit
$g\rightarrow -\infty$ one has $\varepsilon_{\text{CM}}
\rightarrow \varepsilon_0$ and the curves are expected to merge.

For moderate attraction
the behavior of the black solid curve in Fig.~\ref{omega0_632} 
strongly departs from those of the dashed and dotted curves. In fact,
$1/\gamma_p$ interpolates between
dashed and dotted curves, first showing a minimum as
$\left|g\right|$ is decreased and then going to $+\infty$ as
$g\rightarrow 0$. Such trend complies with the physical
expectation that correlated tunneling is forbidden in the absence 
of interaction and that it is favored by the extension of the pair, 
the larger the pair size the fatter the wave function tail in the
barrier.  

\subsection{Discussion}\label{s:discussion}

In principle pair tunneling may be investigated numerically 
simulating the full quantum mechanical time evolution 
of two atoms that are allowed to
escape from the trap, as it was done for repulsive
interactions in Refs.~\onlinecite{Lode2009,Kim2011b,Lode2012,Hunn2013}.
However, the present case of attractive interactions raises 
a computational issue on the accuracy of the interacting wave function.
In fact, the two-body wave function in the relative frame collapses
in space with increasing attraction \cite{Busch1998}. 
Hence, a larger basis set is neeeded in typical
variational methods 
\cite{Rontani2006,Lode2009,Kim2011b,Lode2012,Hunn2013}
to provide a certain numerical accuracy,
which implies either a higher-energy cutoff or a higher resolution in
real space, as we discuss at length elsewhere \cite{DAmico2013}.

Besides, it is difficult to treat numerically realistic tunnel
barriers that are typically shallow, as the one shown in Fig.~\ref{frame}.
As a matter of fact, previous numerical approaches for repulsive 
interactions \cite{Lode2009,Kim2011b,Lode2012,Hunn2013}
considered only idealized functional forms of potential profiles. 
The approximate theory presented in this work is 
fit to any potential profile and interaction strength.

\section{Comparison with the Heidelberg experiment}\label{s:result}

Since the tunneling rates in the experiment
\cite{Zuern2013} are the outcome of a complex fitting
procedure involving different measurements, for the sake of
clarity we consider a single observable, that is the decay time of $P_2(t)$. 
Such quantity is easily obtained in both theory and experiment. 
In the former it is simply $1/(\gamma_s + \gamma_p)$ 
according to Eq.~\eqref{eq:p2t}, whereas
in the latter it is a straightforward exponential fitting to
measured data, as shown in Fig.~\ref{WKB} for the noninteracting case. 

\begin{figure}
\begin{center}
\includegraphics[width=8.0cm, angle=0]{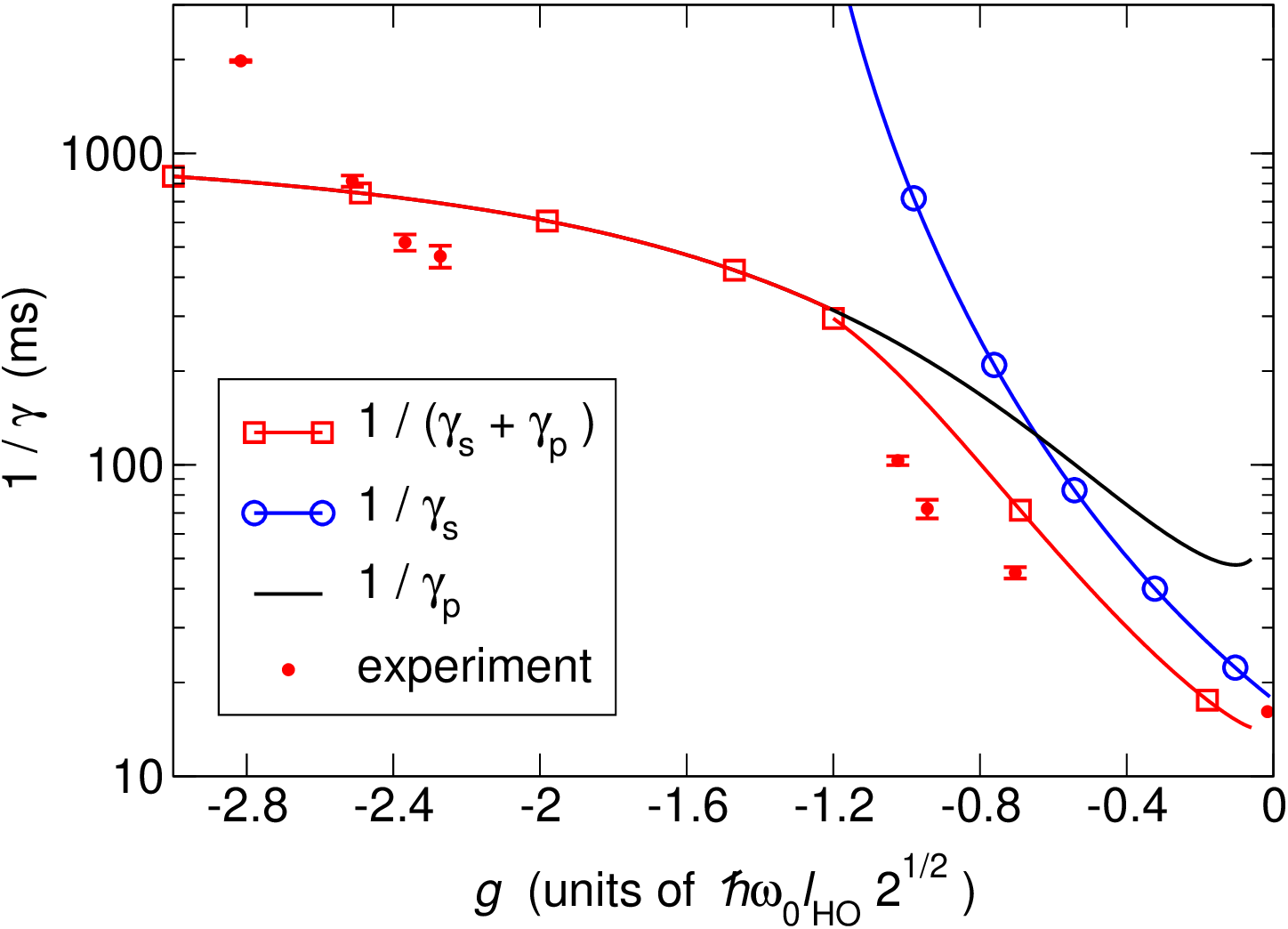}
\end{center}
\caption{ 
(color online).
Inverse tunneling rates $1/\gamma$ vs coupling constant $g$
for $\omega_0 = 250$ Hz $\times$ $2\pi$.
$g$ is in units of $\hbar\omega_0(2\hbar/m\omega_0)^{1/2}$.
The points with their error bars
are the measured values of $1/(\gamma_s+\gamma_p)$,
extracted from Table III of Ref.~\onlinecite{Zuern2013} 
(column $\gamma_{2,\text{fit}}$) except for the rightmost point
that is taken from our Fig.~\ref{WKB}.
The red [light gray]
curve is the theoretical estimate, as explained in the main text.
\label{taus_omega0_25}}
\end{figure}

Figure \ref{taus_omega0_25} compares the measured and predicted 
values of $1/(\gamma_s + \gamma_p)$ as a function
of the coupling constant $g$.
We remark that in our theory the fitting parameter is the HO frequency
$\omega_0$. This fixes the interaction energy for a certain value of 
$g$ \cite{Busch1998}, whereas
in Ref.~\onlinecite{Zuern2013} the interaction energy
is the output of the fitting procedure, the fitting parameters
being the tunneling rates. Besides, the determination of $\omega_0$
is a non-trivial experimental task, being specific to the type
of spectroscopy \cite{Zuern2013,Wenz2013}.

The natural choice for the free parameter $\omega_0$
would be $\omega_0=2\omega_{\text{WKB}}$,
as one may regards $\hbar\omega_{\text{WKB}}$ 
as the zero-point energy of the HO.
This was also the value chosen to compute the tunneling rates 
in the interacting case shown in Fig.~\ref{omega0_632}.
However, the predicted values of the tunneling rates
turn out to be systematically 
small with respect to the measured values \cite{Zuern2013}.

The chosen value of $\omega_0$ in Fig.~\ref{taus_omega0_25} 
is $\omega_0=$ 250 Hz $\times$ $2\pi$, with
$\varepsilon_0=\hbar\omega_{\text{WKB}}$ and
$E_{\text{trap}}(g)=\hbar\omega_0\left[\tilde{E}_{\text{trap}}(g)-1/2
\right] +\hbar\omega_{\text{WKB}}$, $\tilde{E}_{\text{trap}}(g)$
being the relative-motion energy in units of $\hbar\omega_0$
according to Ref.~\onlinecite{Busch1998}.  
The blue [gray] curve is the quasiparticle prediction for $1/\gamma_s$,
which deviates only slightly (within 10\%) from the 
noninteracting WKB value $1/(2\gamma_{s0})$. 
The black curve is $1/\gamma_p$ calculated 
following the method outlined in subsection \ref{s:PT}.
The points are the measured values of $1/(\gamma_s+\gamma_p)$
with their error bars, taken from Table III of Ref.~\onlinecite{Zuern2013}
(column $\gamma_{2,\text{fit}}$) except for the rightmost point
that is taken from our Fig.~\ref{WKB}.
Note that the values of $g$ were rescaled due to the
different HO reference frequency. 

The theoretical value of $1/(\gamma_s+\gamma_p)$  
(solid red [light gray] curve) 
fairly compares with the measured points, exhibiting the same 
qualitative trend. We note that at small values of $\left|g\right|$
the measured values of $1/(\gamma_s+\gamma_p)$
are smaller than the theoretical ones, whereas at large $\left|g\right|$
the opposite holds. This suggests that the effective 
frequency $\omega_0$ increases with $\left|g\right|$, since 
the smaller $\omega_0$ the smaller $1/(\gamma_s+\gamma_p)$. 
Such behavior appears reasonable
as the HO frequency obtained by expanding $V(x)$
around the trap bottom is larger than $2\omega_{\text{WKB}}$.
Therefore, one expects larger anharmonic effects at higher energies, i.e.,
at smaller values of $\left|g\right|$. This confirms
\emph{a posteriori} that neglecting the effects of the anharmonic terms
of the potential on the two-body wave function is a reasonable approximation. 

All data measured in Ref.~\onlinecite{Zuern2013} were explained 
assuming no pair tunneling, $\gamma_p=0$. However, the error of
the measurements performed in the regime of strong attraction was
too large to exclude unambiguously the occurrence of pair tunneling. 
Figure \ref{taus_omega0_25} shows that the channel associated
to usual single-atom tunneling (blue [gray] solid line) closes already
at moderate values of $g\sim 1$. This prediction paves the way to
future experiments in the regime of moderate attraction, where  
only pair tunneling is expected to survive.

\section{Conclusions}

In this Article we have developed a theory of the
pair tunneling of two fermions out of a trap that is
based on simple and physically transparent formulae. 
We predict that the
observation of pair tunneling is within reach
of present experiments with $^6$Li atoms. 
Intriguingly, it was recently showed \cite{Rontani2009c,Zuern2013} 
that pairing emerges already with very few atoms in 
tight low-dimensional traps.  
Therefore, pair tunneling may provide an important spectroscopic tool
to address pairing in many-body states. 

\begin{acknowledgments}
This work is supported by the EU-FP7 Marie Curie initial training 
network INDEX and by the CINECA-ISCRA grants IscrC\_TUN1DFEW 
and IscrC\_TRAP-DIP.
I thank Gerhard Z\"urn and Selim Jochim for stimulating discussions
as well as for sharing their results prior to publication, making
available the numerical data.
\end{acknowledgments}

%\bibliography{pair}

\begin{thebibliography}{10}%
\makeatletter
\providecommand \@ifxundefined [1]{%
 \ifx #1\undefined \expandafter \@firstoftwo
 \else \expandafter \@secondoftwo
\fi
}%
\providecommand \@ifnum [1]{%
 \ifnum #1\expandafter \@firstoftwo
 \else \expandafter \@secondoftwo
\fi
}%
\providecommand \enquote [1]{``#1''}%
\providecommand \bibnamefont  [1]{#1}%
\providecommand \bibfnamefont [1]{#1}%
\providecommand \citenamefont [1]{#1}%
\providecommand\href[0]{\@sanitize\@href}%
\providecommand\@href[1]{\endgroup\@@startlink{#1}\endgroup\@@href}%
\providecommand\@@href[1]{#1\@@endlink}%
\providecommand \@sanitize [0]{\begingroup\catcode`\&12\catcode`\#12\relax}%
\@ifxundefined \pdfoutput {\@firstoftwo}{%
 \@ifnum{\z@=\pdfoutput}{\@firstoftwo}{\@secondoftwo}%
}{%
 \providecommand\@@startlink[1]{\leavevmode}%
 \providecommand\@@endlink[0]{}%
}{%
 \providecommand\@@startlink[1]{%
  \leavevmode
  \pdfstartlink
   attr{/Border[0 0 1 ]/H/I/C[0 1 1]}%
   user{/Subtype/Link/A<</Type/Action/S/URI/URI(#1)>>}%
  \relax
 }%
 \providecommand\@@endlink[0]{\pdfendlink}%
}%
\providecommand \url  [0]{\begingroup\@sanitize \@url }%
\providecommand \@url [1]{\endgroup\@href {#1}{\urlprefix}}%
\providecommand \urlprefix [0]{URL }%
\providecommand \Eprint[0]{\href }%
\@ifxundefined \urlstyle {%
  \providecommand \doi [1]{doi:\discretionary{}{}{}#1}%
}{%
  \providecommand \doi [0]{doi:\discretionary{}{}{}\begingroup
  \urlstyle{rm}\Url }%
}%
\providecommand \doibase [0]{http://dx.doi.org/}%
\providecommand \Doi[1]{\href{\doibase#1}}%
\providecommand \bibAnnote [3]{%
  \BibitemShut{#1}%
  \begin{quotation}\noindent
    \textsc{Key:}\ #2\\\textsc{Annotation:}\ #3%
  \end{quotation}%
}%
\providecommand \bibAnnoteFile [2]{%
  \IfFileExists{#2}{\bibAnnote {#1} {#2} {\input{#2}}}{}%
}%
\providecommand \typeout [0]{\immediate \write \m@ne }%
\providecommand \selectlanguage [0]{\@gobble}%
\providecommand \bibinfo [0]{\@secondoftwo}%
\providecommand \bibfield [0]{\@secondoftwo}%
\providecommand \translation [1]{[#1]}%
\providecommand \BibitemOpen[0]{}%
\providecommand \bibitemStop [0]{}%
\providecommand \bibitemNoStop [0]{.\EOS\space}%
\providecommand \EOS [0]{\spacefactor3000\relax}%
\providecommand \BibitemShut [1]{\csname bibitem#1\endcsname}%
%</preamble>
\bibitem{deGennes1999}%
  \BibitemOpen
  \bibfield{author}{%
  \bibinfo {author} {\bibfnamefont{P.~G.}\ \bibnamefont{de~Gennes}},\ }%
  \emph{\bibinfo {title} {Superconductivity of metals and alloys}}\ (\bibinfo
  {publisher} {Westview Press},\ \bibinfo {address} {Boulder (Colorado)},\
  \bibinfo {year} {1999})%
  \bibAnnoteFile{NoStop}{deGennes1999}%
\bibitem{Migdal1960}%
  \BibitemOpen
  \bibfield{author}{%
  \bibinfo {author} {\bibfnamefont{A.~B.}\ \bibnamefont{Migdal}},\ }%
  \bibfield{journal}{%
  \bibinfo {journal} {Zh. Eksp. i Teor. Fiz. Pisma}\ }%
  \textbf{\bibinfo {volume} {37}},\ \bibinfo {pages} {249} (\bibinfo {year}
  {1959}),\ \bibinfo {note} {[Sov. Phys.--JETP {\bf 10,} 176 (1960)]}%
  \bibAnnoteFile{NoStop}{Migdal1960}%
\bibitem{Bohr2008}%
  \BibitemOpen
  \bibfield{author}{%
  \bibinfo {author} {\bibfnamefont{A.}~\bibnamefont{Bohr}}\ and\ \bibinfo
  {author} {\bibfnamefont{B.~R.}\ \bibnamefont{Mottelson}},\ }%
  \emph{\bibinfo {title} {Nuclear structure --- Vol. I and II}}\ (\bibinfo
  {publisher} {World Scientific},\ \bibinfo {address} {Singapore},\ \bibinfo
  {year} {1998})%
  \bibAnnoteFile{NoStop}{Bohr2008}%
\bibitem{Ginzburg1964}%
  \BibitemOpen
  \bibfield{author}{%
  \bibinfo {author} {\bibfnamefont{V.~L.}\ \bibnamefont{Ginzburg}}\ and\
  \bibinfo {author} {\bibfnamefont{D.~A.}\ \bibnamefont{Kirzhnits}},\ }%
  \bibfield{journal}{%
  \bibinfo {journal} {Zh. Eksp. i Teor. Fiz. Pisma}\ }%
  \textbf{\bibinfo {volume} {47}},\ \bibinfo {pages} {2006} (\bibinfo {year}
  {1965}),\ \bibinfo {note} {[Sov. Phys.--JETP {\bf 20,} 1346 (1965)]}%
  \bibAnnoteFile{NoStop}{Ginzburg1964}%
\bibitem{Pines1969}%
  \BibitemOpen
  \bibfield{author}{%
  \bibinfo {author} {\bibfnamefont{D.}~\bibnamefont{Pines}}, \bibinfo {author}
  {\bibfnamefont{G.}~\bibnamefont{Baym}},\ and\ \bibinfo {author}
  {\bibfnamefont{C.}~\bibnamefont{Pethick}},\ }%
  \bibfield{journal}{%
  \bibinfo {journal} {Nature}\ }%
  \textbf{\bibinfo {volume} {224}},\ \bibinfo {pages} {673} (\bibinfo {year}
  {1969})%
  \bibAnnoteFile{NoStop}{Pines1969}%
\bibitem{Legget2006}%
  \BibitemOpen
  \bibfield{author}{%
  \bibinfo {author} {\bibfnamefont{A.~J.}\ \bibnamefont{Leggett}},\ }%
  \emph{\bibinfo {title} {Quantum liquids}},\ \bibinfo {edition} {1st}\ ed.\
  (\bibinfo {publisher} {Oxford University Press},\ \bibinfo {address}
  {Oxford},\ \bibinfo {year} {2006})%
  \bibAnnoteFile{NoStop}{Legget2006}%
\bibitem{Rontani2013}%
  \BibitemOpen
  \bibfield{author}{%
  \bibinfo {author} {\bibfnamefont{M.}~\bibnamefont{Rontani}}\ and\ \bibinfo
  {author} {\bibfnamefont{L.~J.}\ \bibnamefont{Sham}},\ }%
  \enquote{\bibinfo {title} {Novel superfluids volume 2},}\ \ (\bibinfo
  {publisher} {Oxford University Press},\ \bibinfo {address} {Oxford, UK},\
  \bibinfo {year} {2013})\ \bibinfo {note} {preprint at arXiv:1301.1726}%
  \bibAnnoteFile{NoStop}{Rontani2013}%
\bibitem{BCS1957}%
  \BibitemOpen
  \bibfield{author}{%
  \bibinfo {author} {\bibfnamefont{J.}~\bibnamefont{Bardeen}}, \bibinfo
  {author} {\bibfnamefont{L.~N.}\ \bibnamefont{Cooper}},\ and\ \bibinfo
  {author} {\bibfnamefont{J.~R.}\ \bibnamefont{Schrieffer}},\ }%
  \bibfield{journal}{%
  \bibinfo {journal} {Phys. Rev.}\ }%
  \textbf{\bibinfo {volume} {108}},\ \bibinfo {pages} {1175} (\bibinfo {year}
  {1957})%
  \bibAnnoteFile{NoStop}{BCS1957}%
\bibitem{Barone1982}%
  \BibitemOpen
  \bibfield{author}{%
  \bibinfo {author} {\bibfnamefont{A.}~\bibnamefont{Barone}}\ and\ \bibinfo
  {author} {\bibfnamefont{G.}~\bibnamefont{Paterno}},\ }%
  \emph{\bibinfo {title} {Physics and applications of the Josephson effect}}\
  (\bibinfo {publisher} {Wiley},\ \bibinfo {address} {New York},\ \bibinfo
  {year} {1982})%
  \bibAnnoteFile{NoStop}{Barone1982}%
\bibitem{Andreev1964}%
  \BibitemOpen
  \bibfield{author}{%
  \bibinfo {author} {\bibfnamefont{A.~F.}\ \bibnamefont{Andreev}},\ }%
  \bibfield{journal}{%
  \bibinfo {journal} {Zh. Eksp. i Teor. Fiz.}\ }%
  \textbf{\bibinfo {volume} {46}},\ \bibinfo {pages} {1823} (\bibinfo {year}
  {1964}),\ \bibinfo {note} {[Sov. Phys.--JETP {\bf 19,} 1228 (1964)]}%
  \bibAnnoteFile{NoStop}{Andreev1964}%
\bibitem{Serwane2011}%
  \BibitemOpen
  \bibfield{author}{%
  \bibinfo {author} {\bibfnamefont{F.}~\bibnamefont{Serwane}}, \bibinfo
  {author} {\bibfnamefont{G.}~\bibnamefont{Z{\"u}rn}}, \bibinfo {author}
  {\bibfnamefont{T.}~\bibnamefont{Lompe}}, \bibinfo {author}
  {\bibfnamefont{T.~B.}\ \bibnamefont{Ottenstein}}, \bibinfo {author}
  {\bibfnamefont{A.~N.}\ \bibnamefont{Wenz}},\ and\ \bibinfo {author}
  {\bibfnamefont{S.}~\bibnamefont{Jochim}},\ }%
  \bibfield{journal}{%
  \bibinfo {journal} {Science}\ }%
  \textbf{\bibinfo {volume} {332}},\ \bibinfo {pages} {336} (\bibinfo {year}
  {2011})%
  \bibAnnoteFile{NoStop}{Serwane2011}%
\bibitem{Zuern2012}%
  \BibitemOpen
  \bibfield{author}{%
  \bibinfo {author} {\bibfnamefont{G.}~\bibnamefont{Z{\"u}rn}}, \bibinfo
  {author} {\bibfnamefont{F.}~\bibnamefont{Serwane}}, \bibinfo {author}
  {\bibfnamefont{T.}~\bibnamefont{Lompe}}, \bibinfo {author}
  {\bibfnamefont{A.~N.}\ \bibnamefont{Wenz}}, \bibinfo {author}
  {\bibfnamefont{M.~G.}\ \bibnamefont{Ries}}, \bibinfo {author}
  {\bibfnamefont{J.~E.}\ \bibnamefont{Bohn}},\ and\ \bibinfo {author}
  {\bibfnamefont{S.}~\bibnamefont{Jochim}},\ }%
  \bibfield{journal}{%
  \bibinfo {journal} {Phys. Rev. Lett.}\ }%
  \textbf{\bibinfo {volume} {108}},\ \bibinfo {pages} {075303} (\bibinfo {year}
  {2012})%
  \bibAnnoteFile{NoStop}{Zuern2012}%
\bibitem{Zuern2013}%
  \BibitemOpen
  \bibfield{author}{%
  \bibinfo {author} {\bibfnamefont{G.}~\bibnamefont{Z{\"u}rn}}, \bibinfo
  {author} {\bibfnamefont{A.~N.}\ \bibnamefont{Wenz}}, \bibinfo {author}
  {\bibfnamefont{S.}~\bibnamefont{Murmann}}, \bibinfo {author}
  {\bibfnamefont{T.}~\bibnamefont{Lompe}},\ and\ \bibinfo {author}
  {\bibfnamefont{S.}~\bibnamefont{Jochim}}}%
   (\bibinfo {year} {2013}),\ \bibinfo {note} {arXiv:1307.5153v1}%
  \bibAnnoteFile{NoStop}{Zuern2013}%
\bibitem{Chin2010}%
  \BibitemOpen
  \bibfield{author}{%
  \bibinfo {author} {\bibfnamefont{C.}~\bibnamefont{Chin}}, \bibinfo {author}
  {\bibfnamefont{R.}~\bibnamefont{Grimm}}, \bibinfo {author}
  {\bibfnamefont{P.}~\bibnamefont{Julienne}},\ and\ \bibinfo {author}
  {\bibfnamefont{E.}~\bibnamefont{Tiesinga}},\ }%
  \bibfield{journal}{%
  \bibinfo {journal} {Rev. Mod. Phys.}\ }%
  \textbf{\bibinfo {volume} {82}},\ \bibinfo {pages} {1225} (\bibinfo {year}
  {2010})%
  \bibAnnoteFile{NoStop}{Chin2010}%
\bibitem{Regal2003}%
  \BibitemOpen
  \bibfield{author}{%
  \bibinfo {author} {\bibfnamefont{C.~A.}\ \bibnamefont{Regal}}, \bibinfo
  {author} {\bibfnamefont{C.}~\bibnamefont{Ticknor}}, \bibinfo {author}
  {\bibfnamefont{J.~L.}\ \bibnamefont{Bohn}},\ and\ \bibinfo {author}
  {\bibfnamefont{D.~S.}\ \bibnamefont{Jin}},\ }%
  \bibfield{journal}{%
  \bibinfo {journal} {Nature (London)}\ }%
  \textbf{\bibinfo {volume} {424}},\ \bibinfo {pages} {47} (\bibinfo {year}
  {2003})%
  \bibAnnoteFile{NoStop}{Regal2003}%
\bibitem{Bartenstein2004}%
  \BibitemOpen
  \bibfield{author}{%
  \bibinfo {author} {\bibfnamefont{M.}~\bibnamefont{Bartenstein}}, \bibinfo
  {author} {\bibfnamefont{A.}~\bibnamefont{Altmeyer}}, \bibinfo {author}
  {\bibfnamefont{S.}~\bibnamefont{Riedl}}, \bibinfo {author}
  {\bibfnamefont{S.}~\bibnamefont{Jochim}}, \bibinfo {author}
  {\bibfnamefont{C.}~\bibnamefont{Chin}}, \bibinfo {author}
  {\bibfnamefont{J.~H.}\ \bibnamefont{Denschlag}},\ and\ \bibinfo {author}
  {\bibfnamefont{R.}~\bibnamefont{Grimm}},\ }%
  \bibfield{journal}{%
  \bibinfo {journal} {Phys. Rev. Lett.}\ }%
  \textbf{\bibinfo {volume} {92}},\ \bibinfo {pages} {120401} (\bibinfo {year}
  {2004})%
  \bibAnnoteFile{NoStop}{Bartenstein2004}%
\bibitem{Zwierlein2005}%
  \BibitemOpen
  \bibfield{author}{%
  \bibinfo {author} {\bibfnamefont{M.~W.}\ \bibnamefont{Zwierlein}}, \bibinfo
  {author} {\bibfnamefont{J.~R.}\ \bibnamefont{Abo-Shaeer}}, \bibinfo {author}
  {\bibfnamefont{A.}~\bibnamefont{Schirotzek}}, \bibinfo {author}
  {\bibfnamefont{C.~H.}\ \bibnamefont{Schunck}},\ and\ \bibinfo {author}
  {\bibfnamefont{W.}~\bibnamefont{Ketterle}},\ }%
  \bibfield{journal}{%
  \bibinfo {journal} {Nature (London)}\ }%
  \textbf{\bibinfo {volume} {435}},\ \bibinfo {pages} {1047} (\bibinfo {year}
  {2005})%
  \bibAnnoteFile{NoStop}{Zwierlein2005}%
\bibitem{Giorgini2008}%
  \BibitemOpen
  \bibfield{author}{%
  \bibinfo {author} {\bibfnamefont{S.}~\bibnamefont{Giorgini}}, \bibinfo
  {author} {\bibfnamefont{L.~P.}\ \bibnamefont{Pitaevskii}},\ and\ \bibinfo
  {author} {\bibfnamefont{S.}~\bibnamefont{Stringari}},\ }%
  \bibfield{journal}{%
  \bibinfo {journal} {Rev. Mod. Phys.}\ }%
  \textbf{\bibinfo {volume} {80}},\ \bibinfo {pages} {1215} (\bibinfo {year}
  {2008})%
  \bibAnnoteFile{NoStop}{Giorgini2008}%
\bibitem{Rontani2012}%
  \BibitemOpen
  \bibfield{author}{%
  \bibinfo {author} {\bibfnamefont{M.}~\bibnamefont{Rontani}},\ }%
  \bibfield{journal}{%
  \bibinfo {journal} {Phys. Rev. Lett.}\ }%
  \textbf{\bibinfo {volume} {108}},\ \bibinfo {pages} {115302} (\bibinfo {year}
  {2012})%
  \bibAnnoteFile{NoStop}{Rontani2012}%
\bibitem{Dudarev2007}%
  \BibitemOpen
  \bibfield{author}{%
  \bibinfo {author} {\bibfnamefont{A.~M.}\ \bibnamefont{Dudarev}}, \bibinfo
  {author} {\bibfnamefont{M.~G.}\ \bibnamefont{Raizen}},\ and\ \bibinfo
  {author} {\bibfnamefont{Q.}~\bibnamefont{Niu}},\ }%
  \bibfield{journal}{%
  \bibinfo {journal} {Phys. Rev. Lett.}\ }%
  \textbf{\bibinfo {volume} {98}},\ \bibinfo {pages} {063001} (\bibinfo {year}
  {2007})%
  \bibAnnoteFile{NoStop}{Dudarev2007}%
\bibitem{Cheinet2008}%
  \BibitemOpen
  \bibfield{author}{%
  \bibinfo {author} {\bibfnamefont{P.}~\bibnamefont{Cheinet}}, \bibinfo
  {author} {\bibfnamefont{S.}~\bibnamefont{Trotzky}}, \bibinfo {author}
  {\bibfnamefont{M.}~\bibnamefont{Feld}}, \bibinfo {author}
  {\bibfnamefont{U.}~\bibnamefont{Schnorrberger}}, \bibinfo {author}
  {\bibfnamefont{M.}~\bibnamefont{Moreno-Cardoner}}, \bibinfo {author}
  {\bibfnamefont{S.}~\bibnamefont{F{\"o}lling}},\ and\ \bibinfo {author}
  {\bibfnamefont{I.}~\bibnamefont{Bloch}},\ }%
  \bibfield{journal}{%
  \bibinfo {journal} {Phys. Rev. Lett.}\ }%
  \textbf{\bibinfo {volume} {101}},\ \bibinfo {pages} {090404} (\bibinfo {year}
  {2008})%
  \bibAnnoteFile{NoStop}{Cheinet2008}%
\bibitem{delCampo2006}%
  \BibitemOpen
  \bibfield{author}{%
  \bibinfo {author} {\bibfnamefont{A.}~\bibnamefont{del Campo}}, \bibinfo
  {author} {\bibfnamefont{F.}~\bibnamefont{Delgado}}, \bibinfo {author}
  {\bibfnamefont{G.}~\bibnamefont{Gar{c\'{\i}}a-Calder{\'o}n}}, \bibinfo
  {author} {\bibfnamefont{J.~G.}\ \bibnamefont{Muga}},\ and\ \bibinfo {author}
  {\bibfnamefont{M.~G.}\ \bibnamefont{Raizen}},\ }%
  \bibfield{journal}{%
  \bibinfo {journal} {Phys. Rev. A}\ }%
  \textbf{\bibinfo {volume} {74}},\ \bibinfo {pages} {013605} (\bibinfo {year}
  {2006})%
  \bibAnnoteFile{NoStop}{delCampo2006}%
\bibitem{delCampo2011}%
  \BibitemOpen
  \bibfield{author}{%
  \bibinfo {author} {\bibfnamefont{A.}~\bibnamefont{del Campo}},\ }%
  \bibfield{journal}{%
  \bibinfo {journal} {Phys. Rev. A}\ }%
  \textbf{\bibinfo {volume} {84}},\ \bibinfo {pages} {012113} (\bibinfo {year}
  {2011})%
  \bibAnnoteFile{NoStop}{delCampo2011}%
\bibitem{Calderon2011}%
  \BibitemOpen
  \bibfield{author}{%
  \bibinfo {author}
  {\bibfnamefont{G.}~\bibnamefont{Gar{c\'{\i}}a-Calder{\'o}n}}\ and\ \bibinfo
  {author} {\bibfnamefont{L.~G.}\ \bibnamefont{Mendoza-Luna}},\ }%
  \bibfield{journal}{%
  \bibinfo {journal} {Phys. Rev. A}\ }%
  \textbf{\bibinfo {volume} {84}},\ \bibinfo {pages} {032106} (\bibinfo {year}
  {2011})%
  \bibAnnoteFile{NoStop}{Calderon2011}%
\bibitem{Sokolovski2012}%
  \BibitemOpen
  \bibfield{author}{%
  \bibinfo {author} {\bibfnamefont{D.}~\bibnamefont{Sokolovski}}, \bibinfo
  {author} {\bibfnamefont{M.}~\bibnamefont{Pons}},\ and\ \bibinfo {author}
  {\bibfnamefont{T.}~\bibnamefont{Kamalov}},\ }%
  \bibfield{journal}{%
  \bibinfo {journal} {Phys. Rev. A}\ }%
  \textbf{\bibinfo {volume} {86}},\ \bibinfo {pages} {022110} (\bibinfo {year}
  {2012})%
  \bibAnnoteFile{NoStop}{Sokolovski2012}%
\bibitem{Longhi2012}%
  \BibitemOpen
  \bibfield{author}{%
  \bibinfo {author} {\bibfnamefont{S.}~\bibnamefont{Longhi}}\ and\ \bibinfo
  {author} {\bibfnamefont{G.~D.}\ \bibnamefont{Valle}},\ }%
  \bibfield{journal}{%
  \bibinfo {journal} {Phys. Rev. A}\ }%
  \textbf{\bibinfo {volume} {86}},\ \bibinfo {pages} {012112} (\bibinfo {year}
  {2012})%
  \bibAnnoteFile{NoStop}{Longhi2012}%
\bibitem{Pons2012}%
  \BibitemOpen
  \bibfield{author}{%
  \bibinfo {author} {\bibfnamefont{M.}~\bibnamefont{Pons}}, \bibinfo {author}
  {\bibfnamefont{D.}~\bibnamefont{Sokolovski}},\ and\ \bibinfo {author}
  {\bibfnamefont{A.}~\bibnamefont{del Campo}},\ }%
  \bibfield{journal}{%
  \bibinfo {journal} {Phys. Rev. A}\ }%
  \textbf{\bibinfo {volume} {85}},\ \bibinfo {pages} {022107} (\bibinfo {year}
  {2012})%
  \bibAnnoteFile{NoStop}{Pons2012}%
\bibitem{Georgiou2012}%
  \BibitemOpen
  \bibfield{author}{%
  \bibinfo {author} {\bibfnamefont{O.}~\bibnamefont{Georgiou}}, \bibinfo
  {author} {\bibfnamefont{G.}~\bibnamefont{Glicori{\'c}}}, \bibinfo {author}
  {\bibfnamefont{A.}~\bibnamefont{Lazarides}}, \bibinfo {author}
  {\bibfnamefont{D.~F.~M.}\ \bibnamefont{Oliveira}}, \bibinfo {author}
  {\bibfnamefont{J.~D.}\ \bibnamefont{Bodyfelt}},\ and\ \bibinfo {author}
  {\bibfnamefont{A.}~\bibnamefont{Goussev}},\ }%
  \bibfield{journal}{%
  \bibinfo {journal} {Europhys. Lett.}\ }%
  \textbf{\bibinfo {volume} {100}},\ \bibinfo {pages} {20005} (\bibinfo {year}
  {2012})%
  \bibAnnoteFile{NoStop}{Georgiou2012}%
\bibitem{Girardeau1960}%
  \BibitemOpen
  \bibfield{author}{%
  \bibinfo {author} {\bibfnamefont{M.}~\bibnamefont{Girardeau}},\ }%
  \bibfield{journal}{%
  \bibinfo {journal} {J. Math. Phys. (N.Y.)}\ }%
  \textbf{\bibinfo {volume} {1}},\ \bibinfo {pages} {516} (\bibinfo {year}
  {1960})%
  \bibAnnoteFile{NoStop}{Girardeau1960}%
\bibitem{Cheon1999}%
  \BibitemOpen
  \bibfield{author}{%
  \bibinfo {author} {\bibfnamefont{T.}~\bibnamefont{Cheon}}\ and\ \bibinfo
  {author} {\bibfnamefont{T.}~\bibnamefont{Shigehara}},\ }%
  \bibfield{journal}{%
  \bibinfo {journal} {Phys. Rev. Lett.}\ }%
  \textbf{\bibinfo {volume} {82}},\ \bibinfo {pages} {2536} (\bibinfo {year}
  {1999})%
  \bibAnnoteFile{NoStop}{Cheon1999}%
\bibitem{Zollner2008}%
  \BibitemOpen
  \bibfield{author}{%
  \bibinfo {author} {\bibfnamefont{S.}~\bibnamefont{Z{\"o}llner}}, \bibinfo
  {author} {\bibfnamefont{H.-D.}\ \bibnamefont{Meyer}},\ and\ \bibinfo {author}
  {\bibfnamefont{P.}~\bibnamefont{Schmelcher}},\ }%
  \bibfield{journal}{%
  \bibinfo {journal} {Phys. Rev. Lett.}\ }%
  \textbf{\bibinfo {volume} {100}},\ \bibinfo {pages} {040401} (\bibinfo {year}
  {2008})%
  \bibAnnoteFile{NoStop}{Zollner2008}%
\bibitem{Streltsov2011}%
  \BibitemOpen
  \bibfield{author}{%
  \bibinfo {author} {\bibfnamefont{A.~I.}\ \bibnamefont{Streltsov}}, \bibinfo
  {author} {\bibfnamefont{K.}~\bibnamefont{Sakmann}}, \bibinfo {author}
  {\bibfnamefont{O.~E.}\ \bibnamefont{Alon}},\ and\ \bibinfo {author}
  {\bibfnamefont{L.~S.}\ \bibnamefont{Cederbaum}},\ }%
  \bibfield{journal}{%
  \bibinfo {journal} {Phys. Rev. A}\ }%
  \textbf{\bibinfo {volume} {83}},\ \bibinfo {pages} {043604} (\bibinfo {year}
  {2011})%
  \bibAnnoteFile{NoStop}{Streltsov2011}%
\bibitem{Chatterjee2012}%
  \BibitemOpen
  \bibfield{author}{%
  \bibinfo {author} {\bibfnamefont{B.}~\bibnamefont{Chatterjee}}, \bibinfo
  {author} {\bibfnamefont{I.}~\bibnamefont{Brouzos}}, \bibinfo {author}
  {\bibfnamefont{L.}~\bibnamefont{Cao}},\ and\ \bibinfo {author}
  {\bibfnamefont{P.}~\bibnamefont{Schmelcher}},\ }%
  \bibfield{journal}{%
  \bibinfo {journal} {Phys. Rev. A}\ }%
  \textbf{\bibinfo {volume} {85}},\ \bibinfo {pages} {013611} (\bibinfo {year}
  {2012})%
  \bibAnnoteFile{NoStop}{Chatterjee2012}%
\bibitem{Hunn2013}%
  \BibitemOpen
  \bibfield{author}{%
  \bibinfo {author} {\bibfnamefont{S.}~\bibnamefont{Hunn}}, \bibinfo {author}
  {\bibfnamefont{K.}~\bibnamefont{Zimmermann}}, \bibinfo {author}
  {\bibfnamefont{M.}~\bibnamefont{Hiller}},\ and\ \bibinfo {author}
  {\bibfnamefont{A.}~\bibnamefont{Buchleitner}},\ }%
  \bibfield{journal}{%
  \bibinfo {journal} {Phys. Rev. A}\ }%
  \textbf{\bibinfo {volume} {87}},\ \bibinfo {pages} {043626} (\bibinfo {year}
  {2013})%
  \bibAnnoteFile{NoStop}{Hunn2013}%
\bibitem{Bugnion2013}%
  \BibitemOpen
  \bibfield{author}{%
  \bibinfo {author} {\bibfnamefont{P.~O.}\ \bibnamefont{Bugnion}}\ and\
  \bibinfo {author} {\bibfnamefont{G.~J.}\ \bibnamefont{Conduit}},\ }%
  \bibfield{journal}{%
  \bibinfo {journal} {Phys. Rev. A}\ }%
  \textbf{\bibinfo {volume} {88}},\ \bibinfo {pages} {013601} (\bibinfo {year}
  {2013})%
  \bibAnnoteFile{NoStop}{Bugnion2013}%
\bibitem{Lode2009}%
  \BibitemOpen
  \bibfield{author}{%
  \bibinfo {author} {\bibfnamefont{A.~U.~J.}\ \bibnamefont{Lode}}, \bibinfo
  {author} {\bibfnamefont{A.~I.}\ \bibnamefont{Streltsov}}, \bibinfo {author}
  {\bibfnamefont{O.~E.}\ \bibnamefont{Alon}}, \bibinfo {author}
  {\bibfnamefont{H.-D.}\ \bibnamefont{Meyer}},\ and\ \bibinfo {author}
  {\bibfnamefont{L.~S.}\ \bibnamefont{Cederbaum}},\ }%
  \bibfield{journal}{%
  \bibinfo {journal} {J. Phys. B}\ }%
  \textbf{\bibinfo {volume} {42}},\ \bibinfo {pages} {044018} (\bibinfo {year}
  {2009})%
  \bibAnnoteFile{NoStop}{Lode2009}%
\bibitem{Kim2011b}%
  \BibitemOpen
  \bibfield{author}{%
  \bibinfo {author} {\bibfnamefont{S.}~\bibnamefont{Kim}}\ and\ \bibinfo
  {author} {\bibfnamefont{J.}~\bibnamefont{Brand}},\ }%
  \bibfield{journal}{%
  \bibinfo {journal} {J. Phys. B: At. Mol. Opt. Phys.}\ }%
  \textbf{\bibinfo {volume} {44}},\ \bibinfo {pages} {195301} (\bibinfo {year}
  {2011})%
  \bibAnnoteFile{NoStop}{Kim2011b}%
\bibitem{Lode2012}%
  \BibitemOpen
  \bibfield{author}{%
  \bibinfo {author} {\bibfnamefont{A.~U.~J.}\ \bibnamefont{Lode}}, \bibinfo
  {author} {\bibfnamefont{A.~I.}\ \bibnamefont{Streltsov}}, \bibinfo {author}
  {\bibfnamefont{K.}~\bibnamefont{Sakmann}}, \bibinfo {author}
  {\bibfnamefont{O.~E.}\ \bibnamefont{Alon}},\ and\ \bibinfo {author}
  {\bibfnamefont{L.~S.}\ \bibnamefont{Cederbaum}},\ }%
  \bibfield{journal}{%
  \bibinfo {journal} {Proc. Natl. Acad. Sci. USA}\ }%
  \textbf{\bibinfo {volume} {109}},\ \bibinfo {pages} {13521} (\bibinfo {year}
  {2012})%
  \bibAnnoteFile{NoStop}{Lode2012}%
\bibitem{Bardeen1961}%
  \BibitemOpen
  \bibfield{author}{%
  \bibinfo {author} {\bibfnamefont{J.}~\bibnamefont{Bardeen}},\ }%
  \bibfield{journal}{%
  \bibinfo {journal} {Phys. Rev. Lett.}\ }%
  \textbf{\bibinfo {volume} {6}},\ \bibinfo {pages} {57} (\bibinfo {year}
  {1961})%
  \bibAnnoteFile{NoStop}{Bardeen1961}%
\bibitem{Girardeau2010}%
  \BibitemOpen
  \bibfield{author}{%
  \bibinfo {author} {\bibfnamefont{M.~D.}\ \bibnamefont{Girardeau}}\ and\
  \bibinfo {author} {\bibfnamefont{G.~E.}\ \bibnamefont{Astrakharchik}},\ }%
  \bibfield{journal}{%
  \bibinfo {journal} {Phys. Rev. A}\ }%
  \textbf{\bibinfo {volume} {81}},\ \bibinfo {pages} {061601(R)} (\bibinfo
  {year} {2010})%
  \bibAnnoteFile{NoStop}{Girardeau2010}%
\bibitem{Girardeau2010b}%
  \BibitemOpen
  \bibfield{author}{%
  \bibinfo {author} {\bibfnamefont{M.~D.}\ \bibnamefont{Girardeau}},\ }%
  \bibfield{journal}{%
  \bibinfo {journal} {Phys. Rev. A}\ }%
  \textbf{\bibinfo {volume} {82}},\ \bibinfo {pages} {011607(R)} (\bibinfo
  {year} {2010})%
  \bibAnnoteFile{NoStop}{Girardeau2010b}%
\bibitem{Bugnion2013b}%
  \BibitemOpen
  \bibfield{author}{%
  \bibinfo {author} {\bibfnamefont{P.~O.}\ \bibnamefont{Bugnion}}\ and\
  \bibinfo {author} {\bibfnamefont{G.~J.}\ \bibnamefont{Conduit}},\ }%
  \bibfield{journal}{%
  \bibinfo {journal} {Phys. Rev. A}\ }%
  \textbf{\bibinfo {volume} {87}},\ \bibinfo {pages} {060502(R)} (\bibinfo
  {year} {2013})%
  \bibAnnoteFile{NoStop}{Bugnion2013b}%
\bibitem{Taniguchi2011}%
  \BibitemOpen
  \bibfield{author}{%
  \bibinfo {author} {\bibfnamefont{T.}~\bibnamefont{Taniguchi}}\ and\ \bibinfo
  {author} {\bibfnamefont{S.~I.}\ \bibnamefont{Sawada}},\ }%
  \bibfield{journal}{%
  \bibinfo {journal} {Phys. Rev. E}\ }%
  \textbf{\bibinfo {volume} {83}},\ \bibinfo {pages} {026208} (\bibinfo {year}
  {2011})%
  \bibAnnoteFile{NoStop}{Taniguchi2011}%
\bibitem{Razavy2003}%
  \BibitemOpen
  \bibfield{author}{%
  \bibinfo {author} {\bibfnamefont{M.}~\bibnamefont{Razavy}},\ }%
  \emph{\bibinfo {title} {Quantum theory of tunneling}}\ (\bibinfo {publisher}
  {World Scientific},\ \bibinfo {address} {Singapore},\ \bibinfo {year}
  {2003})%
  \bibAnnoteFile{NoStop}{Razavy2003}%
\bibitem{Wilkinson1997}%
  \BibitemOpen
  \bibfield{author}{%
  \bibinfo {author} {\bibfnamefont{S.~R.}\ \bibnamefont{Wilkinson}}, \bibinfo
  {author} {\bibfnamefont{C.~F.}\ \bibnamefont{Bharucha}}, \bibinfo {author}
  {\bibfnamefont{M.~C.}\ \bibnamefont{Fisher}}, \bibinfo {author}
  {\bibfnamefont{K.~W.}\ \bibnamefont{Madison}}, \bibinfo {author}
  {\bibfnamefont{P.~R.}\ \bibnamefont{Morrow}}, \bibinfo {author}
  {\bibfnamefont{Q.}~\bibnamefont{Niu}}, \bibinfo {author}
  {\bibfnamefont{B.}~\bibnamefont{Sundaram}},\ and\ \bibinfo {author}
  {\bibfnamefont{M.~G.}\ \bibnamefont{Raizen}},\ }%
  \bibfield{journal}{%
  \bibinfo {journal} {Nature (London)}\ }%
  \textbf{\bibinfo {volume} {387}},\ \bibinfo {pages} {575} (\bibinfo {year}
  {1997})%
  \bibAnnoteFile{NoStop}{Wilkinson1997}%
\bibitem{Rothe2006}%
  \BibitemOpen
  \bibfield{author}{%
  \bibinfo {author} {\bibfnamefont{C.}~\bibnamefont{Rothe}}, \bibinfo {author}
  {\bibfnamefont{S.~I.}\ \bibnamefont{Hintschich}},\ and\ \bibinfo {author}
  {\bibfnamefont{A.~P.}\ \bibnamefont{Monkman}},\ }%
  \bibfield{journal}{%
  \bibinfo {journal} {Phys. Rev. Lett.}\ }%
  \textbf{\bibinfo {volume} {96}},\ \bibinfo {pages} {163601} (\bibinfo {year}
  {2006})%
  \bibAnnoteFile{NoStop}{Rothe2006}%
\bibitem{Wang2012b}%
  \BibitemOpen
  \bibfield{author}{%
  \bibinfo {author} {\bibfnamefont{J.-J.}\ \bibnamefont{Wang}}, \bibinfo
  {author} {\bibfnamefont{W.}~\bibnamefont{Li}}, \bibinfo {author}
  {\bibfnamefont{S.}~\bibnamefont{Chen}}, \bibinfo {author}
  {\bibfnamefont{G.}~\bibnamefont{Xianlong}}, \bibinfo {author}
  {\bibfnamefont{M.}~\bibnamefont{Rontani}},\ and\ \bibinfo {author}
  {\bibfnamefont{M.}~\bibnamefont{Polini}},\ }%
  \bibfield{journal}{%
  \bibinfo {journal} {Phys. Rev. B}\ }%
  \textbf{\bibinfo {volume} {86}},\ \bibinfo {pages} {075110} (\bibinfo {year}
  {2012})%
  \bibAnnoteFile{NoStop}{Wang2012b}%
\bibitem{Busch1998}%
  \BibitemOpen
  \bibfield{author}{%
  \bibinfo {author} {\bibfnamefont{T.}~\bibnamefont{Busch}}, \bibinfo {author}
  {\bibfnamefont{B.}~\bibnamefont{Englert}}, \bibinfo {author}
  {\bibfnamefont{K.}~\bibnamefont{Rz{\c{a}\.z}ewski}},\ and\ \bibinfo {author}
  {\bibfnamefont{M.}~\bibnamefont{Wilkens}},\ }%
  \bibfield{journal}{%
  \bibinfo {journal} {Found. Phys.}\ }%
  \textbf{\bibinfo {volume} {28}},\ \bibinfo {pages} {549} (\bibinfo {year}
  {1998})%
  \bibAnnoteFile{NoStop}{Busch1998}%
\bibitem{Rontani2006}%
  \BibitemOpen
  \bibfield{author}{%
  \bibinfo {author} {\bibfnamefont{M.}~\bibnamefont{Rontani}}, \bibinfo
  {author} {\bibfnamefont{C.}~\bibnamefont{Cavazzoni}}, \bibinfo {author}
  {\bibfnamefont{D.}~\bibnamefont{Bellucci}},\ and\ \bibinfo {author}
  {\bibfnamefont{G.}~\bibnamefont{Goldoni}},\ }%
  \bibfield{journal}{%
  \bibinfo {journal} {J. Chem. Phys.}\ }%
  \textbf{\bibinfo {volume} {124}},\ \bibinfo {pages} {124102} (\bibinfo {year}
  {2006})%
  \bibAnnoteFile{NoStop}{Rontani2006}%
\bibitem{DAmico2013}%
  \BibitemOpen
  \bibfield{author}{%
  \bibinfo {author} {\bibfnamefont{P.}~\bibnamefont{D'Amico}}\ and\ \bibinfo
  {author} {\bibfnamefont{M.}~\bibnamefont{Rontani}}}%
   (\bibinfo {year} {2013}),\ \bibinfo {note} {arXiv:1310.3829}%
  \bibAnnoteFile{NoStop}{DAmico2013}%
\bibitem{Wenz2013}%
  \BibitemOpen
  \bibfield{author}{%
  \bibinfo {author} {\bibfnamefont{A.~N.}\ \bibnamefont{Wenz}}, \bibinfo
  {author} {\bibfnamefont{G.}~\bibnamefont{Z{\"u}rn}}, \bibinfo {author}
  {\bibfnamefont{S.}~\bibnamefont{Murmann}}, \bibinfo {author}
  {\bibfnamefont{I.}~\bibnamefont{Brouzos}}, \bibinfo {author}
  {\bibfnamefont{T.}~\bibnamefont{Lompe}},\ and\ \bibinfo {author}
  {\bibfnamefont{S.}~\bibnamefont{Jochim}}}%
   (\bibinfo {year} {2013}),\ \bibinfo {note} {arXiv:1307.3443v1}%
  \bibAnnoteFile{NoStop}{Wenz2013}%
\bibitem{Rontani2009c}%
  \BibitemOpen
  \bibfield{author}{%
  \bibinfo {author} {\bibfnamefont{M.}~\bibnamefont{Rontani}}, \bibinfo
  {author} {\bibfnamefont{J.~R.}\ \bibnamefont{Armstrong}}, \bibinfo {author}
  {\bibfnamefont{Y.}~\bibnamefont{Yu}}, \bibinfo {author}
  {\bibfnamefont{S.}~\bibnamefont{{\AA}berg}},\ and\ \bibinfo {author}
  {\bibfnamefont{S.~M.}\ \bibnamefont{Reimann}},\ }%
  \bibfield{journal}{%
  \bibinfo {journal} {Phys. Rev. Lett.}\ }%
  \textbf{\bibinfo {volume} {102}},\ \bibinfo {pages} {060401} (\bibinfo {year}
  {2009})%
  \bibAnnoteFile{NoStop}{Rontani2009c}%
\end{thebibliography}

%Merlin.mbs v4.21 2009-07-09.
%

\end{document}